\documentclass[11pt,a4paper]{article}

\usepackage{amsmath,amssymb}     
\usepackage{color}
\usepackage{graphicx}
\usepackage{subfig}
\usepackage{cite}                
\usepackage{hyperref}            
\usepackage{multirow,makecell}
\usepackage{braket,bm}
\usepackage{cancel}
\RequirePackage{doi}
\usepackage{hyperref}
\usepackage{leftidx}
\usepackage[text={17cm,24.5cm},centering]{geometry}
\usepackage{textcomp}
\usepackage{wasysym}

\numberwithin{equation}{section}   

\def \be {\begin{equation}}
\def \ee {\end{equation}}
\def \ba {\begin{array}}
\def \ea {\end{array}}
\def \bea{\begin{eqnarray}}
\def \eea{\end{eqnarray}}

\def \a {\alpha}
\def \b {\beta}
\def \g {\gamma}
\def \G {\Gamma}
\def \d {\delta}
\def \D {\Delta}
\def \dg {\dagger}
\def \e {\epsilon}

\def \m {\mu}

\def \l {\lambda}
\def \L {\Lambda}
\def \vp {\varphi}
\def \s {\sigma}

\def \r {\rho}

\def \O {\Omega}

\def \p {\partial}

\def \mc {\mathcal}

\def \inf {\infty}

\def \tr {\textrm{tr}}

\def \and {{\textrm{and}}}

\begin{document}
\begin{titlepage}
	
	\title{\textbf {Symmetry decomposition of relative entropies in conformal field theory}}
	\author{Hui-Huang Chen\footnote{chenhh@jxnu.edu.cn}~,}
	\date{}
	
	\maketitle
	\underline{}
	\vspace{-12mm}
	
	\begin{center}
		{\it
             College of Physics and Communication Electronics, Jiangxi Normal University,\\ Nanchang 330022, China\\
		}
		\vspace{10mm}
	\end{center}
	\begin{abstract}
	 We consider the symmetry resolution of relative entropies in the 1+1 dimensional free massless compact boson conformal field theory (CFT) which presents an internal $U(1)$ symmetry. We calculate various symmetry resolved R\'enyi relative entropies between one interval reduced density matrices of CFT primary states using the replica method. By taking the replica limit, the symmetry resolved relative entropy can be obtained. We also take the XX spin chain model as a concrete lattice realization of this CFT to perform numerical computation. The CFT predictions are tested against exact numerical calculations finding perfect agreement.
	\end{abstract}
	
\end{titlepage}

\thispagestyle{empty}

\newpage

\tableofcontents
\section{Introduction}
In recent years, concepts and methods coming from quantum information theory are playing more and more important roles in both high-energy physics and condensed matter theory.
In a many-body system, entanglement is a powerful tool to characterize quantum phase transition and by studying its universal features one can acquire knowledge of the underlying conformal field theory (CFT). For reviews, see \cite{Amico:2007ag, Calabrese:2009qy, Eisert:2008ur, laflorencie2016quantum}. In high-energy physics, entanglement is also the key concept to understand the information paradox of black holes \cite{Hawking:1974sw, Hawking:1976ra, mathur2009information} through gauge/gravity duality \cite{Maldacena:1997re, Almheiri:2020cfm}.
\par So far, most studies are focus on subsystem entanglement features of a single quantum state. For some applications, the entanglement entropy for a given subsystem can not provide enough information. One may wonder, how can we gain insight when giving two different quantum states. It's also important for us to distinguish between subsystems in different states. In this respect, relative entropy is an important quantity \cite{Vedral:2002zz}. Relative entropy attracts a great deal of attention during the past few years and has been extensively studied \cite{Blanco:2013joa,Lashkari:2014yva,Lashkari:2015dia,Sarosi:2016oks,Ruggiero:2016khg,Jafferis:2015del,Casini:2016udt, Bhattacharya:2012mi, Balakrishnan:2017bjg, Casini:2016fgb}. The reason is that relative entropy is relatively simple to calculate and is free of divergence in quantum field theory. There also exist other quantities that can be used to distinguish reduced density matrices (RDMs). For example, the quantum fidelity \cite{Zhou:2008zza} and the trace distance \cite{Zhang:2019wqo,Zhang:2019itb} is two very commonly used concepts.
\par For two given states with reduced density matrices (RDMs) $\rho$ and $\sigma$, the so-called relative entropy is defined as \cite{Kullback:1951zyt, Araki:1976zv}
\be
S(\rho\|\sigma)=\tr(\rho\log\rho)-\tr(\rho\log\sigma),
\ee
which can be viewed as a measure of ``distance" between the two quantum states.
In quantum field theory, the relative entropy can be obtained by using the replica trick \cite{Lashkari:2014yva, Lashkari:2015dia}
\be
S(\rho\|\sigma)=\lim_{n\rightarrow 1}S_n(\rho\|\s)=\lim_{n\rightarrow 1}\frac{1}{1-n}\log{\frac{\tr(\rho\sigma^{n-1})}{\tr(\rho^n)}},
\ee
where we have defined the R\'enyi relative entropies as
\be\label{Snrhosigma}
S_n(\rho\|\s)=\frac{1}{1-n}\log{\frac{\tr(\rho\sigma^{n-1})}{\tr(\rho^n)}}.
\ee
\par For a quantum many-body systems with global symmetry, one can decompose entanglement into different symmetry sectors. In this respect, the authors of reference \cite{Goldstein:2017bua} introduced a more refined notion of entanglement, the symmetry resolved entanglement entropy. After this pioneering work, people have studied a lot about symmetry resolution of entanglement properties for both pure states\cite{Bonsignori:2019naz, Murciano:2020vgh, Bonsignori:2020laa, Fraenkel:2019ykl, Estienne:2020txv, Azses:2020wfx, Vitale:2021lds}  and mixed states \cite{Cornfeld:2018wbg, Murciano:2021djk}. Moreover, similar quantities have also been introduced in quantum field theories and in the holographic settings \cite{Caputa:2013eka,Belin:2013uta,Belin:2014mva,Caputa:2015qbk,Zhao:2020qmn, Caputa:2015tua, Dowker:2015dpd, Dowker:2016ugd}.
\par In this paper, we will mainly focus on the symmetry resolution of relative entropies in CFT. More explicitly, we will consider the $U(1)$ symmetry decomposition of relative entropy in free massless compact boson CFT using the twist operator method. We will also check our universal CFT predictions numerically in the XX spin chain.
\par The remaining part of this paper is organized as follows. In section \ref{section2}, we briefly review the CFT approach to the R\'enyi relative entropies between the RDMs of two primary excited states. In section \ref{section3}, we discuss how relative entropies are distributed in different charge sectors. In this section, we define all needed concepts concerning symmetry resolved relative entropy and summarise the known results of the symmetry resolved entanglement entropy which will be useful in the following sections. In section \ref{section4}, we calculate various symmetry resolved relative entropies between primary states in free compact boson CFT. The CFT results are tested in section \ref{section5} against exact numerical computations in the XX chain. Finally, we conclude in section \ref{section6} and some technical details for numerical calculation are given in appendix \ref{appenA}.
\section{Relative entopy in CFT}\label{section2}
In this section, let's briefly review the replica trick to compute the relative entropies of two reduced density matrices of excited states in 1+1 dimensional CFT.
Consider a system with one spatial dimension and a bipartition into two complementary regions $A$ and $\bar{A}$. We take subsystem $A$ given by the interval $[u,v]$ with length $l=v-u$ and $\bar{A}$ is its complement with length $L-l$. Here $L$ is the total length of our periodic 1D system. Given two (pure) states $\ket{\Psi},\ket{\Phi}\in\mathcal{H}=\mathcal{H}_A\otimes\mathcal{H}_{\bar{A}}$, the reduced density matrices of subsystem $A$ is defined by tracing over the points not in $A$.
\be
\rho_{A,\Psi}=\tr_{\bar{A}}\ket{\Psi}\bra{\Psi},\quad \rho_{A,\Phi}=\tr_{\bar{A}}\ket{\Phi}\bra{\Phi}.
\ee
The world sheet of the 1+1 dimensional CFT is an infinite cylinder with circumference $L$ which can be parameterized by introducing the complex coordinate $w=x+i\tau$. In this paper, we are only interested in the excited states in CFT that correspond to local primary operators
\be
\ket{\Psi}=\Psi(-i\inf)\ket{0},
\ee
where $\ket{0}$ is the CFT vacuum state and corresponds to the identity operator $I$. Let us omit the index $A$ and denote the reduced density matrix of a state $\ket{\Psi}$ to the subsystem $A$ by $\rho_{\Psi}$. Following the standard procedure \cite{Calabrese:2004eu,Calabrese:2009qy}, $\tr(\rho_{I}^n)$ can be obtained by sewing cyclically $n$ copies of the above cylinders along with the interval $[u,v]$. In contrast to the ground state case, the corresponding path-integral representation of the density matrix $\rho=\ket{\Psi}\bra{\Psi}$ presents two additional insertions of $\Psi(-i\inf)$ and $\Psi^{\dg}(i\inf)$. In this way, we end up with a $n$-sheeted Riemann surface $\mc{R}_n$ and $\tr(\rho_{\Psi}^n)$ is given by a $2n$-point function on $\mc{R}_n$ \cite{Berganza:2011mh}
\be
\tr(\rho_{\Psi}^n)=\frac{Z_n}{Z_1^n}\frac{\langle\prod_{k=1}^n\Psi(w^-_k)\Psi^{\dg}(w^+_k)\rangle_{\mc{R}_n}}{\langle \Psi(w^-_1)\Psi^{\dg}(w_1^+)\rangle^n_{\mc{R}_1}},
\ee
where $Z_n=\langle I\rangle_{\mc{R}_n}$ is the $n$-th moment of the reduced density matrix of the ground state and $w_k^-=-i\inf,w_k^+=i\inf$ are points where the operators are inserted in the $k$-th copy.
\par In order to obtain the R\'enyi relative entropies between $\rho_{\Psi}$ and $\rho_{\Phi}$, we further need to compute $\tr(\rho_{\Psi}\rho_{\Phi}^{n-1})$. Quite similar to the previous case, and taking the normalization factor into account, we find
\be
\tr(\rho_{\Psi}\rho_{\Phi}^{n-1})=\frac{Z_n}{Z_1^n}\frac{\langle \Psi(w_1^-)\Psi^{\dg}(w^+_1)\prod_{k=2}^{n}\Phi(w^-_k)\Phi^{\dg}(w^+_k)\rangle_{\mc{R}_n}}{\langle \Psi(w_1^-)\Psi^{\dg}(w_1^+)\rangle_{\mc{R}_1}\langle \Phi(w_1^-)\Phi^{\dg}(w_1^+)\rangle^{n-1}_{\mc{R}_1}},
\ee
and the universal ratio
\be
G_n(\rho_{\Psi}\|\rho_{\Phi})\equiv\frac{\tr(\rho_{\Psi}\rho_{\Phi}^{n-1})}{\tr(\rho_{\Psi}^n)}=\frac{\langle \Psi(w^-_1)\Psi^{\dg}(w^+_1)\prod_{k=2}^{n}\Phi(w^-_k)\Phi^{\dg}(w^+_k)\rangle_{\mc{R}_n}\langle \Psi(w_1^-)\Psi^{\dg}(w_1^+)\rangle^{n-1}_{\mc{R}_1}}{\langle\prod_{k=1}^n\Psi(w^-_k)\Psi^{\dg}(w^+_k)\rangle_{\mc{R}_n}\langle \Phi(w^-_1)\Phi^{\dg}(w_1^+)\rangle^{n-1}_{\mc{R}_1}}.
\ee
Knowing $G_n$, the R\'enyi relative entropy is simply given by
\be
S_n(\rho_{\Psi}\|\rho_{\Phi})=\frac{1}{1-n}\log G_n(\rho_{\Psi}\|\rho_{\Phi}).
\ee
We can apply the following sequence of conformal maps
\be\label{confmap}
w\rightarrow z=\left(\frac{\sin\frac{\pi(w-u)}{L}}{\sin\frac{\pi(w-u)}{L}}\right)^{\frac{1}{n}}\rightarrow t=-i\log z
\ee
to transform the $n$-sheet Riemann surface $\mc{R}_n$ into a single cylinder. The transformation law of a primary field $\mc{O}$  is very simple
\be
\mc{O}(z,\bar{z})=\left(\frac{dz}{dw}\right)^{-h_{\mc{O}}}\left(\frac{d\bar{z}}{d\bar{w}}\right)^{-\bar{h}_{\mc{O}}}\mc{O}(w,\bar{w}),
\ee
with $(h_{\mc{O}},\bar{h}_{\mc{O}})$ the conformal weights of $\mc{O}$. Applying the conformal maps in eq.~(\ref{confmap}), one can easily express $G_n(\rho_{\Psi}\|\rho_{\Phi})$ in terms of correlation functions on the cylinder
\be
G_n(\rho_{\Psi}\|\rho_{\Phi})=n^{2(n-1)(h_{\Psi}+\bar{h}_{\Psi}-h_{\Phi}-\bar{h}_{\Phi})}\frac{\langle \Psi(t^-_1)\Psi^{\dg}(t^+_1)\prod_{k=2}^{n}\Phi(t^-_k)\Phi^{\dg}(t^+_k)\rangle_{cy}\langle \Psi(t_1^-)\Psi^{\dg}(t_1^+)\rangle^{n-1}_{cy}}{\langle\prod_{k=1}^n\Psi(t^-_k)\Psi^{\dg}(t^+_k)\rangle_{cy}\langle \Phi(t^-_1)\Phi^{\dg}(t_1^+)\rangle^{n-1}_{cy}},
\ee
where $t_k^{\pm}$ are the points corresponding to $w_k^{\pm}$ through the map $t(w)$
\be\label{tkn}
t_k^-=\frac{\pi}{n}(x+2(k-1)),\qquad t_k^+=\frac{\pi}{n}(-x+2(k-1)),\qquad k=1,2,\cdots,n.\quad x=\frac{v-u}{L}=\frac{l}{L}.
\ee
In the following, we will mainly focus on the theory of free massless compact bosonic field $\vp(z,\bar{z})$, with Euclidean action
\be\label{action}
\mc{A}[\vp]=\frac{1}{8\pi}\int dzd\bar{z}\p_z\varphi\p_{\bar{z}}\vp.
\ee
This is a CFT with central charge $c=1$ and has two types of primary fields. The first type is the vertex operators
\be
V_{\a,\bar{\a}}=:e^{i(\a\phi+\bar{\a}\bar{\phi})}:
\ee
where $\phi,\bar{\phi}$ are chiral and anti-chiral parts of the bosonic field: $\vp(z,\bar{z})=\phi(z)+\bar{\phi}(\bar{z})$. The conformal weight of the vertex operator is
$(h,\bar{h})=(\frac{\a^2}{2},\frac{\bar{\a}^2}{2})$. For simplicity, we will consider holomorphic field $\bar{\a}=0$ only. The $n$-point function of vertex operators on the complex plane is ($z_{i,j}\equiv z_i-z_j$)\cite{DiFrancesco:1997nk}
\be
\langle\prod_{k}V_{\a_k}(z_k)\rangle=\prod_{i<j}(z_{i,j})^{\a_i\a_j}.
\ee
After the conformal map $t=-i\log z$ to the cylinder, this correlator becomes
\be
\langle\prod_{k}V_{\a_k}(t_k)\rangle_{cy}=\prod_{i<j}(2\sin\frac{t_{i,j}}{2})^{\a_i\a_j}.
\ee
The other type of primary field in this theory is the derivative operator $i\p\phi$ with conformal dimension $(h,\bar{h})=(1,0)$. The $2n$-point function on the complex plane is given by \cite{DiFrancesco:1997nk}
\be\label{ipphi}
\langle\prod_{k=1}^{2n}i\p\phi(z_k)\rangle=\text{Hf}\left[\frac{1}{z_{i,j}^2}\right]_{1\leq i,j\leq 2n},
\ee
where $\text{Hf}(A)$ is the Haffnian of the $2n\times 2n$ matrix $A$
\be
\text{Hf}(A)=\frac{1}{2^nn!}\sum_{\s\in S_{2n}}\prod_{i=1}^nA_{\s(2i-1),\s(2i)}.
\ee
The Haffian in eq.~(\ref{ipphi}) can be written as a determinant
\be
\text{Hf}\left[\frac{1}{z_{i,j}^2}\right]_{1\leq i,j\leq 2n}=\det\left[\frac{1}{z_{i,j}}\right]_{1\leq i,j\leq 2n}.
\ee
In a cylinder parametrized by $t=-i\log z$, the correlator becomes
\be
\langle\prod_{k=1}^{2n}i\p\phi(t_k)\rangle_{cy}=\frac{1}{4^n}\det\left[\frac{1}{\sin(t_{i,j}/2)}\right]_{1\leq i,j\leq 2n}.
\ee
For this $2n$-point correlator evaluated at the $2n$-point list in eq.~(\ref{tkn}),
the analytic continuation has been obtained in \cite{2013Shell, 2014Entanglement} and is given by
\be\label{parana}
\langle\prod_{k=1}^ni\p\phi(t_k^-)i\p\phi(t_k^+)\rangle_{cy}=\frac{\Gamma^2(\frac{1+n+n\csc \pi x}{2})}{\Gamma^2(\frac{1-n+n\csc \pi x}{2})}.
\ee
Several relative entropies have been obtained in \cite{Lashkari:2015dia, Sarosi:2016oks}, here we just report the results. Firstly, the R\'enyi relative entropies between the ground state and the vertex operator are given by
\be
S_n(\rho_I\|\rho_{V_{\a}})=S_n(\rho_{V_{\a}}\|\rho_I)=\frac{\a^2}{1-n}\log\frac{\sin\pi x}{n\sin\frac{\pi x}{n}}.
\ee
By taking the replica limit $n\rightarrow 1$, the relative entropy is obtained as
\be\label{SnVI}
S(\rho_I\|\rho_{V_{\a}})=S(\rho_{V_{\a}}\|\rho_I)=\a^2(1-\pi x\cot(\pi x)).
\ee
The relative entropy between two vertex operators is given by
\be
S(\rho_{V_{\a}}\|\rho_{V_{\b}})=S(\rho_{V_{\b}}\|\rho_{V_{\a}})=(\a-\b)^2(1-\pi x\cot(\pi x)).
\ee
The relative entropy between the derivative operator and the ground state is
\be
S(\rho_{i\p\phi}\|\rho_I)=2\log(2\sin(\pi x))+2-2\pi x\cot(\pi x)+2\psi\left(\frac{\csc(\pi x)}{2}\right)+2\sin(\pi x).
\ee
Finally, the relative entropy between the derivative operator and the vertex operator is
\be
S(\rho_{i\p\phi}\|V_{\a})=S(\rho_{i\p\phi}\|\rho_I)+S(\rho_I\|\rho_{V_{\a}}).
\ee
\section{Symmetry resolution of entanglement entropy and relative entropy}\label{section3}
\subsection{Entanglement entropy and relative entropy in charge sectors}
Now assume that the system has an internal $U(1)$ symmetry with conserved charge $Q$. We also take a bipartition of our system into two subsystems, $A$ and its complement $\bar{A}$ as before. When the conserved charge $Q$ is local, it splits as $Q=Q_A+Q_{\bar{A}}$. We further assume that both $\rho$ and $\s$ are eigenstate of $Q$, which imply $[\rho,Q]=0,[\sigma,Q]=0$. Tracing out the degree of freedom in $\bar{A}$, one obtains $[\rho_A,Q_A]=0,[\sigma_A,Q_A]=0 $. Then the density matrix $\rho_A$ and $\sigma_A$ can be written as block diagonal forms, in which each block corresponds to a different charge sector with eigenvalue $q$ of $Q_A$
\be\label{decom}
\rho_A=\oplus_{q}\Pi_q\rho_A=\oplus_{q}p^{\rho}(q)\rho_A(q),\quad \sigma=\oplus_{q}\Pi_q\s_A=\oplus_{q}p^{\s}(q)\s_A(q),
\ee
where $\Pi_q$ is the projector onto the eigenspace of $Q_A$ with fixed eigenvalue $q$. We have
\be\label{rhoAq}
\rho_A(q)=\frac{\Pi_q\rho_A}{\tr(\Pi_q\rho_A)},\quad \s_A(q)=\frac{\Pi_q\s_A}{\tr(\Pi_q\s_A)}.
\ee
The denominators in the above equations are introduced to keep the normalization $\tr\rho_A(q)=1,\tr\s_A(q)=1$, which imply
\be
\tr(\Pi_q\rho_A)=p^{\rho}(q),\quad \tr(\Pi_q\s_A)=p^{\s}(q).
\ee
Here $p^{\rho}(q)$ (or $p^{\s}(q)$, respectively) is the probability of finding $q$ as the outcome of a measurement of $Q_A$ in state $\rho_A$ (resp. $\s_A$).
\par Our goal is to understand how the relative entropy is distributed in different charged sectors. Let's start with the resolution of von Neumann entanglement entropy. The equation (\ref{decom}) implies the following decomposition of entanglement entropy
\be\label{Sq}
S(\rho_A)=\sum_qp^{\rho}(q)S(\rho_A(q))-\sum_{q}p^{\rho}(q)\log p^{\rho}(q)\equiv S^c+S^f,
\ee
where
\be
S(\rho_A(q))=-\tr[\rho_A(q)\log\rho_A(q)].
\ee
is the symmetry resolved entanglement entropy associated to $\rho_A(q)$.
In eq.~(\ref{Sq}), we have divided $S(\rho_A)$ into two parts, $S^c$ and $S^f$, which are called the configurational entanglement entropy and the fluctuation entanglement entropy respectively. The configurational entanglement entropy $S^c=\sum_qp^{\rho}(q)S(\rho_A(q))$, measuring the total entropy of all the charged sectors. The fluctuation entanglement entropy $S^f=-\sum_qp^{\rho}(q)\log p^{\rho}(q)$ takes into account the entropy due to fluctuations of the eigenvalues of the charge.
\par In a similar way, we define the symmetry resolved R\'enyi relative entropies as
\be
S_n(\rho_A(q)\|\s_A(q))=\frac{1}{1-n}\log{\frac{\tr(\rho_A(q)\s_A(q)^{n-1})}{\tr(\rho_A(q)^n)}}.
\ee
After substituting the expression of $\rho_A(q)$ and $\s_A(q)$ given in eq.~(\ref{rhoAq}) into the above equation, we obtain
\bea
S_n(\rho_A(q)\|\s_A(q))=\frac{1}{1-n}\log\frac{p^{\rho}(q)^n}{p^{\s}(q)^{n-1}p^{\rho}(q)}\frac{\tr(\rho_A\s_A^{n-1}\Pi_q)}{\tr(\rho_A^n\Pi_q)}\\
=-\log\frac{p^{\rho}(q)}{p^{\s}(q)}+\frac{1}{1-n}\left[\log\tr(\rho_A\s_A^{n-1}\Pi_q)-\log\tr(\rho_A^n\Pi_q)\right].
\eea
Taking the limit $n\rightarrow 1$, we find
\be
S(\rho_A(q)\|\s_A(q))=-\log\frac{p^{\rho}(q)}{p^{\s}(q)}-\frac{1}{p^{\rho}(q)}\tr(\Pi_q\rho_A\log\s_A)+\frac{1}{p^{\rho}(q)}\tr(\Pi_q\rho_A\log\rho_A).
\ee
Multiplying both sides of the above equation with $p^{\rho}(q)$ and summing over $q$, we get
\be\label{Srhosigma}
S(\rho_A\|\s_A)=\langle S(\rho_A(q)\|\s_A(q))\rangle_{p^{\rho}}+\left\langle\log\frac{p^{\rho}(q)}{p^{\s}(q)}\right\rangle_{p^{\rho}},
\ee
where
\be
\langle S(\rho_A(q)\|\s_A(q))\rangle_{p^{\rho}}=\sum_{q}S(\rho_A(q)\|\s_A(q))p^{\rho}(q),
\ee
is the averaged symmetry resolved relative entropy under the probability distribution $p^{\rho}(q)$, and
\be
\left\langle\log\frac{p^{\rho}(q)}{p^{\s}(q)}\right\rangle_{p^{\rho}}=\sum_{q}p^{\rho}(q)\log\frac{p^{\rho}(q)}{p^{\s}(q)}
\ee
is the classical relative entropy or  Kullback-Leibler divergence of probability distribution $p^{\rho}(q)$ and $p^{\s}(q)$. Here, for the relative entropy, we find the equation (\ref{Srhosigma}) looks quite similar to eq.~(\ref{Sq}). You can call the two terms on the right-hand side of eq.~(\ref{Srhosigma}) the configuration relative entropy and the fluctuation relative entropy respectively if you will.
\par Let's define the following generalized probability distributions
\be\label{pnq}
p_n^{\rho|\s}(q)=\frac{\tr(\rho_A\s_A^{n-1}\Pi_q)}{\tr(\rho_A\s_A^{n-1})},\quad p_n^{\rho}(q)=\frac{\tr(\rho_A^n\Pi_q)}{\tr(\rho_A^n)},
\ee
which are normalized as $\sum_q p_n^{\rho|\s}(q)=\sum_q p_n^{\rho}(q)=1$.
For $n=1$, since $p_1^{\rho|\s}(q)=p_1^{\rho}(q)=p^{\rho}(q)$, these generalized distributions are just the physical probability distribution of the subsystem charge $Q_A$ in the state $\rho_A$.
Using these generalized probabilities, we can rewrite the symmetry resolved R\'enyi relative entropy as
\be\label{Snq}
S_n(\rho_A(q)\|\s_A(q))=-\log\frac{p^{\rho}(q)}{p^{\s}(q)}+\frac{1}{1-n}\log\frac{p_n^{\rho|\s}(q)}{p_n^{\rho}(q)}+S_n(\rho_A\|\s_A).
\ee
Taking the limit $n\rightarrow 1$ of the above equation, we obtain the symmetry resolved relative entropy
\be\label{Srhosigmaq}
S(\rho_A(q)\|\s_A(q))=-\log\frac{p^{\rho}(q)}{p^{\s}(q)}-\frac{1}{p^{\rho}(q)}\partial_n(p_n^{\rho|\s}(q)-p_n^{\rho}(q))\big|_{n=1}+S(\rho_A\|\s_A).
\ee
The average of the above equation over $p^{\rho}(q)$ also gives the equation (\ref{Srhosigma}) using the fact that $\sum_q\partial_n(p_n^{\rho|\s}(q)-p_n^{\rho}(q))\big|_{n=1}=0$. It's also useful to average equation (\ref{Snq}) over $p^{\rho}(q)$ to give another expression of the decomposition of the R\'enyi relative entropy
\be\label{Sn}
S_n(\rho_A\|\s_A)=\langle S_n(\rho_A(q)\|\s_A(q))\rangle_{p^{\rho}}+\left\langle\log\frac{p^{\rho}(q)}{p^{\s}(q)}\right\rangle_{p^{\rho}}+\frac{1}{1-n}\sum_qp^{\rho}(q)\log\frac{p_n^{\rho|\s}(q)}{p_n^{\rho}(q)}.
\ee
\par From eq.~(\ref{Snq}) and eq.~(\ref{Srhosigmaq}), we see that to obtain the symmetry resolved R\'enyi relative entropy and relative entropy, one needs to compute the generalized probabilities $p_n^{\rho|\s}(q)$ and $p_n^{\rho}(q)$.
However, this is very hard in general due to the non-local feature of the projector $\Pi_q$. Similar to the case of computing symmetry resolved entanglement entropy, we can bypass this difficulty by defining the following quantities
\be\label{pnmu}
p_n^{\rho|\s}(\m)=\frac{\tr(\rho_A\s_A^{n-1}e^{i\m Q_A})}{\tr(\rho_A\s_A^{n-1})},\quad p_n^{\rho}(\m)=\frac{\tr(\rho_A^ne^{i\m Q_A})}{\tr(\rho_A^n)},
\ee
which turns out to be much easier to compute.
From eq.~(\ref{pnq}), it's easy to see that $p_n^{\rho|\s}(q)$ and $p_n^{\rho}(q)$ can be obtained by Fourier transformations\footnote{Here we have assumed that the eigenvalues of $Q_A$ are continuous. If the eigenvalues are integers, one needs to change the range of the integral to $[-\pi,\pi]$.}
\be
p_n^{\rho|\s}(q)=\int_{-\inf}^{\inf}\frac{d\m}{2\pi}e^{-i\m q}p_n^{\rho|\s}(\m),\quad p_n^{\rho}(q)=\int_{-\inf}^{\inf}\frac{d\m}{2\pi}e^{-i\m q}p_n^{\rho}(\m).
\ee
Here we have used the same notation but with a different argument to denote the Fourier transform of the generalized probabilities $p_n^{\rho|\s}(q)$ and $p_n^{\rho}(q)$.
\par For future use, it's also useful to define the following ratio
\be
F^{\rho,\s}_n(\m,A)\equiv\frac{p_n^{\rho|\s}(\m)}{p_n^{\rho}(\m)}=\frac{\tr(\rho_A\s_A^{n-1}e^{i\m Q_A})\tr(\rho_A^n)}{\tr(\rho_A^ne^{i\m Q_A})\tr(\rho_A\s_A^{n-1})}.
\ee
\subsection{Symmetry resolution of entanglement entropy in CFT}
From the analysis in the last subsection, we see that to compute the symmetry resolved relative entropy, the first step is the calculation of Fourier transformed generalized probabilities $p_n^{\rho|\s}(\m)$ and $p_n^{\rho}(\m)$.
In the free compact boson CFT, the later ($p_n^{\rho}(\m)$, cf.~(\ref{pnmu})) has already been studied in the context of symmetry resolution of entanglement entropy \cite{Capizzi:2020jed}, where people called it the (normalized) charged moments. In this subsection, we briefly review the results and fix some notations that will be useful in the following sections.
\par Let's first consider the ground state case. When $\rho$ is the ground state of a CFT, $\tr(\rho_A^ne^{i\m Q_A})$ can be seen as a partition function in the $n$-sheet Riemann surface $\mc{R}_n$ with an inserted Aharonov-Bohm flux $\m$. In two-dimensional CFT, the insertion of a flux corresponds to a twisted boundary condition, which can be implemented by some local fields acting on the boundary of subsystem $A$. This operator can be seen as the composition of the branch point twist field $\mc{T}_n$ and the $U(1)$ twist field $\mc{V}_{\m}$ and we denote it by $\mc{T}_{n,\m}$ \cite{Goldstein:2017bua, Cardy:2007mb}. The form factors and vacuum expectation values (VEVs) of the composite twist field in integrable field theories have been obtained in \cite{Horvath:2020vzs, Horvath:2021fks} recently. If subsystem $A$ is an interval $[u,v]$, then one can identify
\be
e^{i\m Q_A}=\mc{V}_{\m}(u,0)\mc{V}_{-\m}(v,0).
\ee
More precisely, we have the following relations
\be\label{Zmu}
\tr(\rho_{I,A}^ne^{i\m Q_A})=\langle e^{i\m Q_A}\rangle_{\mc{R}_n}=\langle\mc{V}_{\m}(u,0)\mc{V}_{-\m}(v,0)\rangle_{\mc{R}_n}=\langle\mc{T}_{n,\m}(u,0)\tilde{\mc{T}}_{n,\m}(v,0)\rangle_{\mc{R}_1},
\ee
where $\tilde{\mc{T}}_{n,\m}$ is the anti-twist field. The conformal weight of $\mc{T}_{n,\m}$ and $\tilde{\mc{T}}_{n,\m}$ are the same and are given by
\be
h_{n,\m}=h_n+\frac{h_{\m}}{n},\qquad h_n=\frac{c}{24}(n-\frac{1}{n}),
\ee
where $c$ is the central charge of the CFT.
Then, on the cylinder of circumference $L$, the two-point function of $\mc{T}_{n,\m}$ in eq.~(\ref{Zmu}) is immediately obtained
\be
\tr(\rho_{I,A}^ne^{i\m Q_A})=s_{n,\m}\left(\frac{L}{\pi}\sin\frac{\pi l}{L}\right)^{-\frac{c}{6}(n-\frac{1}{n})-2\frac{h_{\m}+\bar{h}_{\m}}{n}},
\ee
where $s_{n,\m}$ is the unknown non-universal normalization of the composite twist field.
The charged moments of the ground state are
\be
p^{I}_n(\m)=\frac{\tr(\rho_{I,A}^ne^{i\m Q_A})}{\tr(\rho_{I,A}^n)}=\frac{\langle\mc{T}_{n,\m}(u,0)\tilde{\mc{T}}_{n,\m}(v,0)\rangle_{\mc{R}_1}}{\langle\mc{T}_{n}(u,0)\tilde{\mc{T}}_{n}(v,0)\rangle_{\mc{R}_1}}
=\frac{s_{n,\m}}{s_{n,0}}\left(\frac{L}{\pi}\sin\frac{\pi l}{L}\right)^{-\frac2n(h_{\m}+\bar{h}_{\m})}.
\ee
\par In the free compact bosonic field theory defined by the action eq.~(\ref{action}), the $U(1)$ twist field $\mc{V}_{\m}$ can be implemented by the vertex operator
\be
\mc{V}_{\m}=V_{\frac{\m}{2\pi}}=e^{\frac{i\m}{2\pi}\phi}
\ee
with conformal weight $(h_{\m},\bar{h}_{\m})=(\frac12(\frac{\m}{2\pi})^2,\frac12(\frac{\m}{2\pi})^2)$. Then the charged moments of the ground state or the Fourier transformed generalized probabilities are
\be\label{pnI}
p^{I}_n(\m)\sim\left(\frac{L}{\pi}\sin\frac{\pi l}{L}\right)^{-\frac{\m^2}{2n\pi^2}},
\ee
which are Gaussian distributions and as a consequence, $p^{I}_n(q)$ are also Gaussian distributions. In terms of its variance, we can write $p_n^{I}(q)$ as
\be\label{pnIq}
p_n^{I}(q)=\frac{1}{\sqrt{2\pi\langle\D q^2\rangle_n^I}}\exp(-\frac{\D q^2}{2\langle\D q^2\rangle_n^I}),
\ee
where $\D q^2=(q-\bar{q})^2$ and at large $L$ the variance scales as
\be
\langle\D q^2\rangle_n^I=\int_{-\inf}^{\inf}d q \D q^2p^I_n(q)=\frac{1}{n\pi^2}\log\left(\frac{L}{\pi}\sin\pi x\right)+\g_n+o(1),
\ee
where $\g_n$ is a non-universal constant related to $s_{n,\m}$ and $\bar{q}$ is the mean value of $q$ under the probability distribution $p_n^I(q)$ also cannot be fixed by CFT.\footnote{In the XX spin chain, the exact values of $\g_n$ and $\bar{q}$ have been derived in \cite{Bonsignori:2019naz}.}
Since $p^I_n(\m)$ is Gaussian, we can rewrite it as
\be\label{pnImu}
p^I_n(\m)=\exp\{-\frac12\langle\D q^2\rangle_n^{I}\m^2+i\m\bar{q}\}.
\ee
\par For a excited state $\ket{\Psi}$, it's useful to define the ratio
\be\label{fnPsi}
f_n^{\Psi}(\m,x)\equiv\frac{p^{\Psi}_n(\m)}{p^I_n(\m)}=\frac{\langle\mc{V}_{\m}(i\inf)\mc{V}_{-\m}(-i\inf)\prod_{k=1}^n\Psi(t_k^-)\Psi(t_k^+)\rangle_{cy}}
{\langle\mc{V}_{\m}(i\inf)\mc{V}_{-\m}(-i\inf)\rangle_{cy}\langle\prod_{k=1}^n\Psi(t_k^-)\Psi(t_k^+)\rangle_{cy}}.
\ee
When the excitation is induced by the vertex operator $\Psi=V_{\a}=e^{i\a\phi}$, the computation of eq.~(\ref{fnPsi}) is straightforward and the final result is rather simple \cite{Capizzi:2020jed}
\be
f_n^{V_{\a}}(\m,x)=e^{i\m\a x}.
\ee
Then the charged moments of the vertex operator are
\be
p^{V_{\a}}_n(\m)=f_n^{V_{\a}}(\m,x)p_n^I(\m)=e^{i\m\a x}p_n^I(\m).
\ee
The generalized probability distributions $p^{V_{\a}}_n(q)$ are obtained by Fourier transformation
\be
p^{V_{\a}}_n(q)=p_n^I(q-\a x).
\ee
From the above equation, we conclude that the generalized probability distribution of vertex operators $p^{V_{\a}}_n(q)$ is also Gaussian and having the same variance with the ground state distribution $p^I_n(q)$.
\par However, when the excitation is induced by the derivative operators $\Psi=i\p\phi$, the computation of eq.~(\ref{fnPsi}) is more complicated and we briefly mention the details here. In this calculation, the most involved correlator is
\be
\langle\mc{V}_{\m}(i\inf)\mc{V}_{-\m}(-i\inf)\prod_{k=1}^ni\p\phi(t_k^-)i\p\phi(t_k^+)\rangle_{cy}.
\ee
In the paper \cite{Capizzi:2020jed}, the authors conjectured a formula for this correlator
\be\label{Tparana}
\langle V_{\frac{\m}{2\pi}}(i\inf)V_{-\frac{\m}{2\pi}}(-i\inf)\prod_{k=1}^ni\p\phi(t_k^-)i\p\phi(t_k^+)\rangle_{cy}=
\langle V_{\frac{\m}{2\pi}}(i\inf)V_{-\frac{\m}{2\pi}}(-i\inf)\rangle_{cy}\mc{P}_M(\frac{i\m}{2\pi}),
\ee
where $\mc{P}_M(\l)=\det(M-\l)$ is the characteristic polynomial of the $2n\times 2n$ matrix $M$
\be
M=\frac12\begin{pmatrix}
A&B\\
-B^{T}&A
\end{pmatrix},
\ee
where the matrix elements are ($i,j=1,2,\cdots,n$)
\be
A_{ij}=\begin{cases}
0 & \text{if}\; i=j\\
\frac{1}{\sin[\pi(j-i)/n]}& \text{else}
\end{cases},\qquad
B_{ij}=\frac{1}{\sin[\pi(j-i-x)/n]}.
\ee
\par The analytic continuation of $\mc{P}_M(\l)$ is given by
\be\label{Pana}
\mc{P}_M(\l)=\frac{\G(a_n(\l)+n+1)\G(\bar{a}_n(\l)+n+1)}{\G(a_n(\l)+1)\G(\bar{a}_n(\l)+1)},
\ee
where
\be
a_n(\l)=\frac12(\frac{n}{\sin\pi x}-n-1)+i\l.
\ee
Then after pluging eq.~(\ref{Tparana}), eq.~(\ref{Pana}) and eq.~(\ref{parana}) into eq.~(\ref{fnPsi}), we obtain
\be
f_n^{i\p\phi}(\m,x)=\frac{\G(a_n(\frac{i\m}{2\pi})+n+1)\G(\bar{a}_n(\frac{i\m}{2\pi})+n+1)}{\G(a_n(\frac{i\m}{2\pi})+1)\G(\bar{a}_n(\frac{i\m}{2\pi})+1)}\frac{\G^2(a_n(0)+1)}{\G^2(a_n(0)+n+1)}.
\ee
We can expand $f_n^{i\p\phi}(\m,x)$ in $\m$
\be\label{fn}
\begin{split}
f_n^{i\p\phi}(\m,x)&=\sum_{k=0}^{\inf}\frac{(-1)^k}{(2k)!}b_{n,2k}\m^{2k}
\end{split}
\ee
where $b_{n,2k}$ is the coefficient of $\m^{2k}$ and the first two values are
\be
\begin{split}
b_{n,0}=1,\quad b_{n,2}=\frac{1}{2\pi^2}\left[\psi^{(1)}(a_n(0)+1)-\psi^{(1)}(a_n(0)+n+1)\right].
\end{split}
\ee
Here $\psi^{(n)}(z)$ is the polygamma function.
For integer $n$, the infinite series in eq.~(\ref{fn}) terminate at $k=n$ and as $n\rightarrow 1$, all coefficients except $b_{n,0},b_{n,2}$ vanish. Then we can write
\be
\begin{split}
p_n^{i\p\phi}(q)&=\sum_{k=0}^{\inf}\frac{(-1)^k}{(2k)!}b_{n,2k}\int_{\inf}^{\inf}\frac{d\m}{2\pi}e^{-i\m\D q}\exp\{-\frac12\langle\D q^2\rangle_n^{I}\m^2\}\m^{2k}\\
&=p_n^{I}(q)\sum_{k=0}^{\inf}\frac{b_{n,2k}}{(2k)!}H_{2k}\left(\frac{\D q}{\sqrt{2\langle\D q^2\rangle_n^{I}}}\right)[2\langle\D q^2\rangle_n^{I}]^{-k},
\end{split}
\ee
where $H_{2k}(z)$ is the $2k$-th Hermite polynomial with argument $z$.
\par For $n=1$, we have a very simple result
\be
p^{i\p\phi}(q)=\left(1-\frac{c_1(q)}{\pi^2}\sin^2\pi x\right)p^I(q)=\left(1-\frac{c_1(q)}{\pi^2}\sin^2\pi x\right)\frac{1}{\sqrt{2\pi\langle\D q^2\rangle^I}}\exp(-\frac{\D q^2}{2\langle\D q^2\rangle^I}),
\ee
where $\langle\D q^2\rangle^I\equiv\langle\D q^2\rangle^I_1$ and we have defined
\be
c_n(q)\equiv\frac{\langle\D q^2\rangle_n^{I}-\D q^2}{[\langle\D q^2\rangle_n^{I}]^2}.
\ee
Clearly, $p^{i\p\phi}(q)$ is non-Gaussian.
\section{Symmetry resolution of relative entropy in CFT}\label{section4}
In this section, we will focus on the computation of the symmetry resolved relative entropies between RDMs of excited states induced by primary operators in the free massless compact boson CFT.
According to the analysis in section \ref{section3}, in order to calculate the symmetry resolved relative entropy $S(\rho_{\Psi}(q)\|\rho_{\Phi}(q))$ (see eq.~(\ref{Srhosigmaq})), we have to compute the generalized probability distribution $p_n^{\Psi}(q)$ and $p_n^{\Psi|\Phi}(q)$. Let's first compute the Fourier transform of $p_n^{\Psi}(q)$ and $p_n^{\Psi|\Phi}(q)$
\be\label{ratio1}
p_n^{\Psi}(\m)=\frac{\tr(\rho_{\Psi}^ne^{i\m Q_A})}{\tr(\rho_{\Psi}^n)}=\frac{\langle\mc{V}_{\m}(u_1)\mc{V}_{-\m}(v_1)\prod_{k=1}^n\Psi(w_k^-)\Psi^{\dg}(w_k^+)\rangle_{\mc{R}_n}}
{\langle\prod_{k=1}^n\Psi(w_k^-)\Psi^{\dg}(w_k^+)\rangle_{\mc{R}_n}},
\ee
and
\be\label{pnPsiPhimu}
p_n^{\Psi|\Phi}(\m)=\frac{\tr(\rho_{\Psi}\rho_{\Phi}^{n-1}e^{i\m Q_A})}{\tr(\rho_{\Psi}\rho_{\Phi}^{n-1})}=\frac{\langle\mc{V}_{\m}(u_1)\mc{V}_{-\m}(v_1)\Psi(w_1^-)\Psi^{\dg}(w_1^+)\prod_{k=2}^n\Phi(w_k^-)\Phi^{\dg}(w_k^+)\rangle_{\mc{R}_n}}
{\langle\Psi(w_1^-)\Psi^{\dg}(w_1^+)\prod_{k=2}^n\Phi(w_k^-)\Phi^{\dg}(w_k^+)\rangle_{\mc{R}_n}}.
\ee
Thus their ratio $F^{\Psi,\Phi}_n(\m,x)$ is given by
\be\label{FPsiPhi}
F^{\Psi,\Phi}_n(\m,x)=\frac{\langle\mc{V}_{\m}(u_1)\mc{V}_{-\m}(v_1)\Psi(w_1^-)\Psi^{\dg}(w_1^+)\prod_{k=2}^n\Phi(w_k^-)\Phi^{\dg}(w_k^+)\rangle_{\mc{R}_n}\langle\prod_{k=1}^n\Psi(w_k^-)\Psi^{\dg}(w_k^+)\rangle_{\mc{R}_n}}
{\langle\Psi(w_1^-)\Psi^{\dg}(w_1^+)\prod_{k=2}^n\Phi(w_k^-)\Phi^{\dg}(w_k^+)\rangle_{\mc{R}_n}\langle\mc{V}_{\m}(u_1)\mc{V}_{-\m}(v_1)\prod_{k=1}^n\Psi(w_k^-)\Psi^{\dg}(w_k^+)\rangle_{\mc{R}_n}}.
\ee
One could apply the conformal transformation defined in eq.~(\ref{confmap}) to map the correlators in eq.~(\ref{FPsiPhi}) onto the cylinder. In this mapping, all factors proportional to $\frac{dt}{dw}$ coming from the transformation law of primary fields canceled out, leaving us with
\be\label{FPsiPhi1}
F^{\Psi,\Phi}_n(\m,x)=\frac{\langle\mc{V}_{\m}(i\inf)\mc{V}_{-\m}(-i\inf)\Psi(t_1^-)\Psi^{\dg}(t_1^+)\prod_{k=2}^n\Phi(t_k^-)\Phi^{\dg}(t_k^+)\rangle_{cy}\langle\prod_{k=1}^n\Psi(t_k^-)\Psi^{\dg}(t_k^+)\rangle_{cy}}
{\langle\Psi(t_1^-)\Psi^{\dg}(t_1^+)\prod_{k=2}^n\Phi(t_k^-)\Phi^{\dg}(t_k^+)\rangle_{cy}\langle\mc{V}_{\m}(i\inf)\mc{V}_{-\m}(-i\inf)\prod_{k=1}^n\Psi(t_k^-)\Psi^{\dg}(t_k^+)\rangle_{cy}}.
\ee
If one of the two states is the ground state $\ket{0}$, the above generic formula simplifies to
\be
F^{I,\Phi}_n(\m,x)=\frac{\langle\mc{V}_{\m}(i\inf)\mc{V}_{-\m}(-i\inf)\prod_{k=2}^n\Phi(t_k^-)\Phi^{\dg}(t_k^+)\rangle_{cy}}
{\langle\prod_{k=2}^n\Phi(t_k^-)\Phi^{\dg}(t_k^+)\rangle_{cy}\langle\mc{V}_{\m}(i\inf)\mc{V}_{-\m}(-i\inf)\rangle_{cy}},
\ee
and
\be
F^{\Psi,I}_n(\m,x)=\frac{\langle\mc{V}_{\m}(i\inf)\mc{V}_{-\m}(-i\inf)\Psi(t_1^-)\Psi^{\dg}(t_1^+)\rangle_{cy}\langle\prod_{k=1}^n\Psi(t_k^-)\Psi^{\dg}(t_k^+)\rangle_{cy}}
{\langle\Psi(t_1^-)\Psi^{\dg}(t_1^+)\rangle_{cy}\langle\mc{V}_{\m}(i\inf)\mc{V}_{-\m}(-i\inf)\prod_{k=1}^n\Psi(t_k^-)\Psi^{\dg}(t_k^+)\rangle_{cy}}.
\ee
\subsection{Resolution of relative entropy between the ground state and the vertex operator}
The first case we will consider is the symmetry decomposition of relative entropy between the ground state and the excited state generated by a vertex operator. Firstly, let's compute
\be\label{FIV}
\begin{split}
F^{I,V_{\a}}_n(\m,x)&=\frac{\langle V_{\frac{\m}{2\pi}}(i\inf)V_{-\frac{\m}{2\pi}}(-i\inf)\prod_{k=2}^{n}V_{\a}(t_k^-)V_{-\a}(t_k^+)\rangle_{cy}}
{\langle\prod_{k=2}^{n}V_{\a}(t_k^-)V_{-\a}(t_k^+)\rangle_{cy}\langle V_{\frac{\m}{2\pi}}(i\inf)V_{\frac{\m}{2\pi}}(-i\inf)\rangle_{cy}}\\
&=\prod_{k=2}^{n}\langle V_{\frac{\m}{2\pi}}(i\inf)V_{\a}(t_k^-)\rangle_{cy}\prod_{k=2}^{n}\langle V_{-\frac{\m}{2\pi}}(-i\inf)V_{\a}(t_k^-)\rangle_{cy}\times\\
&\times\prod_{k=2}^{n}\langle V_{\frac{\m}{2\pi}}(i\inf)V_{-\a}(t_k^+)\rangle_{cy}\prod_{k=2}^{n}\langle V_{-\frac{\m}{2\pi}}(-i\inf)V_{-\a}(t_k^+)\rangle_{cy}\\
&=\lim_{\L\rightarrow\inf}\prod_{k=2}^{n}\left(\frac{\sin\frac12(\frac{\pi x+2(k-1)}{n})-i\L}{\sin\frac12(\frac{\pi x+2(k-1)}{n})+i\L}\right)^{\frac{\a\m}{2\pi}}
\prod_{k=2}^{n}\left(\frac{\sin\frac12(\frac{-\pi x+2(k-1)}{n})-i\L}{\sin\frac12(\frac{-\pi x+2(k-1)}{n})+i\L}\right)^{-\frac{\a\m}{2\pi}}\\
&=e^{i\m\a x(1-\frac{1}{n})}.
\end{split}
\ee
Then we find
\be
p^{I|V_{\a}}_n(\m)=F^{I,V_{\a}}_n(\m,x)p^I_n(\m)=e^{i\m\a x(1-\frac{1}{n})}p^I_n(\m).
\ee
Since $p^I_n(\m)$ is a Gaussian distribution, we conclude $p^{I|V_{\a}}_n(q)$ is also a Gaussian with the same variance of $p_n^I(q)$. After Fourier transformation, we find
\be
p^{I|V_{\a}}_n(q)=p_n^I(q-\a x(1-1/n)).
\ee
Then the symmetry resolved R\'enyi relative entropies can be easily derived
\be
\begin{split}
S_n(\rho_I(q)\|\rho_{V_{\a}}(q))=-\log\frac{p^I(q)}{p^{V_{\a}}(q)}+\frac{1}{1-n}\log\frac{p_n^{I|V_{\a}}(q)}{p^{I}_n(q)}+S_n(\rho_I\|\rho_{V_{\a}})\\
=-\log\frac{p^I(q)}{p^{I}(q-\a x)}+\frac{1}{1-n}\log\frac{p_n^{I}(q-(1-1/n)\a x)}{p^{I}_n(q)}+S_n(\rho_I\|\rho_{V_{\a}}).
\end{split}
\ee
After substituting the expression of $p_n^I(q)$ in eq.~(\ref{pnIq}) and R\'enyi relative entropy in eq.~(\ref{SnVI}) into the above equation, we obtain
\be
S_n(\rho_I(q)\|\rho_{V_{\a}}(q))=\frac{2\a x\D q-\a^2x^2}{2\langle\D q^2\rangle^I}-\frac{2n\a x\D q+(1-n)\a^2 x^2}{2n^2\langle\D q^2\rangle_n^I}+\frac{\a^2}{1-n}\log\frac{\sin\pi x}{n\sin\frac{\pi x}{n}}.
\ee
The symmetry resolved relative entropy is obtained by taking the limit $n\rightarrow 1$
\be
S(\rho_I(q)\|\rho_{V_{\a}}(q))=\a^2(1-\pi x\cot(\pi x))-\frac{\a^2x^2}{2\langle\D q^2\rangle^I},
\ee
which is $q$ independent up to order $(\log L)^{-1}$.
\par After a very similar calculation with eq.~(\ref{FIV}), we find
\be
\frac{\langle\prod_{k=1}^{n}V_{\a}(t_k^-)V_{-\a}(t_k^+)\rangle_{cy}\langle V_{\frac{\m}{2\pi}}(i\inf)V_{\frac{\m}{2\pi}}(-i\inf)\rangle_{cy}}{\langle V_{\frac{\m}{2\pi}}(i\inf)V_{-\frac{\m}{2\pi}}(-i\inf)\prod_{k=1}^{n}V_{\a}(t_k^-)V_{-\a}(t_k^+)\rangle_{cy}}=e^{-i\m\a x}.
\ee
Then we have
\be
\begin{split}
F^{V_{\a},I}_n(\m,x)&=\frac{\langle V_{\frac{\m}{2\pi}}(i\inf)V_{-\frac{\m}{2\pi}}(-i\inf)V_{\a}(t_1^-)V_{-\a}(t_1^+)\rangle_{cy}\langle\prod_{k=1}^nV_{\a}(t_k^-)V_{-\a}(t_k^+)\rangle_{cy}}
{\langle V_{\a}(t_1^-)V_{-\a}(t_1^+)\rangle_{cy}\langle V_{\frac{\m}{2\pi}}(i\inf)V_{-\frac{\m}{2\pi}}(-i\inf)\prod_{k=1}^nV_{\a}(t_k^-)V_{-\a}(t_k^+)\rangle_{cy}}\\
&=e^{-i\m\a x}\langle V_{\frac{\m}{2\pi}}(i\inf)V_{\a}(t_1^-)\rangle_{cy}\langle V_{-\frac{\m}{2\pi}}(-i\inf)V_{-\a}(t_1^+)\rangle_{cy}\times\\
&\times\langle V_{\frac{\m}{2\pi}}(i\inf)V_{-\a}(t_1^+)\rangle_{cy}\langle V_{-\frac{\m}{2\pi}}(-i\inf)V_{\a}(t_1^-)\rangle_{cy}\\
&=e^{-i\m\a x}\lim_{\L\rightarrow\inf}\left(-\frac{\sin\frac12(\frac{\pi x}{n}-i\L)}{\sin\frac12(\frac{\pi x}{n}+i\L)}\right)^{\frac{\m\a}{2\pi}}
\left(-\frac{\sin\frac12(\frac{\pi x}{n}-i\L)}{\sin\frac12(\frac{\pi x}{n}+i\L)}\right)^{\frac{\m\a}{2\pi}}\\
&=e^{-i\m\a x(1-\frac{1}{n})}.
\end{split}
\ee
Thus the corresponding Fourier transformed generalized probability distributions can be found as
\be\label{pnVImu}
p^{V_{\a}|I}_n(\m)=F^{V_{\a},I}_n(\m,x)p^{V_{\a}}_n(\m)=F^{V_{\a},I}_n(\m,x)f^{V_{\a}}_n(\m,x)p^I_n(\m)=e^{\frac{i\a\m x}{n}}p^I_n(\m).
\ee
Then we get
\be
p^{V_{\a}|I}_n(q)=p^I_n(q-\a x/n).
\ee
Now the symmetry resolved R\'enyi relative entropies are easily obtained
\be
\begin{split}
S_n(\rho_{V_{\a}}(q)\|\rho_I(q))=-\log\frac{p^{V_{\a}}(q)}{p^I(q)}+\frac{1}{1-n}\log\frac{p_n^{V_{\a}|I}(q)}{p^{V_{\a}}_n(q)}+S_n(\rho_{V_{\a}}\|\rho_I)\\
=\frac{\a^2x^2-2\a x\D q}{2\langle\D q^2\rangle^I}+\frac{2n\a x\D q-(1+n)\a^2 x^2}{2n^2\langle\D q^2\rangle_n^I}+\frac{\a^2}{1-n}\log\frac{\sin\pi x}{n\sin\frac{\pi x}{n}}.
\end{split}
\ee
Taking the limit $n\rightarrow 1$, the symmetry resolved relative entropy is given by
\be
S(\rho_{V_{\a}}(q)\|\rho_I(q))=\a^2(1-\pi x\cot(\pi x))-\frac{\a^2x^2}{2\langle\D q^2\rangle^I}=S(\rho_I(q)\|\rho_{V_{\a}}(q)).
\ee
\par Finally, let's compute the symmetry resolved relative entropy between two vertex operators. As before, we begin with
\be\label{FVaVb}
\begin{split}
F^{V_{\a},V_{\b}}_n(\m,x)&=\frac{\langle\mc{V}_{\m}(i\inf)\mc{V}_{-\m}(-i\inf)V_{\a}(t_1^-)V_{-\a}(t_1^+)\prod_{k=2}^nV_{\b}(t_k^-)V_{-\b}(t_k^+)}
{\langle V_{\a}(t_1^-)V_{-\a}(t_1^+)\prod_{k=2}^nV_{\b}(t_k^-)V_{-\b}(t_k^+)\rangle_{cy}\langle\mc{V}_{\m}(i\inf)\mc{V}_{-\m}(-i\inf)\rangle_{cy}}[f_n^{V_{\a}}(\m,x)]^{-1}\\
&=e^{i\m\a x}(\m,x)F^{V_{\a},I}_n(\m,x)F^{I,V_{\b}}_n(\m,x)[f_n^{V_{\a}}(\m,x)]^{-1}\\
&=e^{i\m(\a-\b)(\frac1n-1)x},
\end{split}
\ee
which implies
\be
p^{V_{\a}|V_{\b}}_n(q)=p_n^I(q-(\a-\b)(1-1/n)x-\a x).
\ee
The symmetry resolved relative entropy can be derived in a similar way, and the final result is
\be
S(\rho_{V_{\a}}(q)\|\rho_{V_{\b}}(q))=(\a-\b)^2(1-\pi x\cot(\pi x))-\frac{(\a-\b)^2x^2}{2\langle\D q^2\rangle^I},
\ee
which is also $q$ independent up to order $1/\log L$.
\par In this subsection, we find that all the symmetry resolved relative entropies did not depend on the charge eigenvalue up to order $(\log L)^{-1}$, which means the relative entropies are the same in the different charge sectors. We must mention that the equipartition of relative entropy may be broken at higher order in $1/\log L$.
\subsection{Resolution of relative entropy between the derivative operator and the ground state}
In this subsection, we will consider a more complicated case, which is the symmetry resolution of the relative entropy between the excited state generated by a derivative operator and the ground state. Let's start with
\be
\begin{split}
F^{i\p\phi,I}_n(\m,x)&=\frac{\langle V_{\frac{\m}{2\pi}}(i\inf)V_{-\frac{\m}{2\pi}}(-i\inf)i\p\phi(t_1^-)i\p\phi(t_1^+)\rangle_{cy}\langle\prod_{k=1}^ni\p\phi(t_k^-)i\p\phi(t_k^+)\rangle_{cy}}
{\langle i\p\phi(t_1^-)i\p\phi(t_1^+)\rangle_{cy}\langle V_{\frac{\m}{2\pi}}(i\inf)V_{-\frac{\m}{2\pi}}(-i\inf)\prod_{k=1}^ni\p\phi(t_k^-)i\p\phi(t_k^+)\rangle_{cy}}\\
&=\frac{\langle V_{\frac{\m}{2\pi}}(i\inf)V_{-\frac{\m}{2\pi}}(-i\inf)i\p\phi(t_1^-)i\p\phi(t_1^+)\rangle_{cy}}
{\langle i\p\phi(t_1^-)i\p\phi(t_1^+)\rangle_{cy}\langle V_{\frac{\m}{2\pi}}(i\inf)V_{-\frac{\m}{2\pi}}(-i\inf)\rangle_{cy}}[f_n^{i\p\phi}(\m,x)]^{-1}\\
&=(1-\frac{\m^2}{\pi^2}\sin^2\frac{\pi x}{n})[f_n^{i\p\phi}(\m,x)]^{-1}.
\end{split}
\ee
Then we obtain
\be\label{pnparImu}
\begin{split}
p^{i\p\phi|I}_n(\m)&=F^{i\p\phi,I}_n(\m,x)p^{i\p\phi}_n(\m)
=F^{i\p\phi,I}_n(\m,x)f_n^{i\p\phi}(\m,x)p_n^I(\m)\\
&=(1-\frac{\m^2}{\pi^2}\sin^2\frac{\pi x}{n})p_n^I(\m).
\end{split}
\ee
After Fourier transformation, we find
\be
p^{i\p\phi|I}_n(q)=\left(1-\frac{c_n(q)}{\pi^2}\sin^2\frac{\pi x}{n}\right)p^I_n(q).
\ee
The symmetry resolved R\'enyi relative entropies can be derived straightforwardly from the following equation
\be
\begin{split}
S_n(\rho_{i\p\phi}(q)\|\rho_{I}(q))=-\log\frac{p^{i\p\phi}(q)}{p^{I}(q)}+\frac{1}{1-n}\log\frac{p_n^{{i\p\phi}|I}(q)}{p^{i\p\phi}_n(q)}+S_n(\rho_{i\p\phi}\|\rho_{I}).\\
\end{split}
\ee
In this case, to see whether equipartition of relative entropy hold at order $1/\log L$, we can keep only the first order $\mu^2$ in the expansion of $f_n^{i\p\phi}(\m,x)$, i.e. to approximate $p_n^{i\p\phi}(q)$ as
\be
p_n^{i\p\phi}(q)\simeq (1-\frac12 b_{n,2}c_n(q))p_n^I(q).
\ee
Then in the physical regime $\D q$ of order 1, we have
\be
S_n(\rho_{i\p\phi}(q)\|\rho_{I}(q))=S_n(\rho_{i\p\phi}\|\rho_{I})+\frac{c_1(q)}{\pi^2}\sin^2\pi x+\frac{c_n(q)}{1-n}(\frac{1}{2}b_{n,2}-\frac{1}{\pi^2}\sin^2\frac{\pi x}{n})+\mathcal{O}((\log L)^{-2}).
\ee
The symmetry resolved relative entropy is obtained after taking the limit $n\rightarrow 1$
\be
\begin{split}
S(\rho_{i\p\phi}(q)\|\rho_{I}(q))=2\log(2\sin(\pi x))+2-2\pi x\cot(\pi x)+2\psi\left(\frac{\csc(\pi x)}{2}\right)+2\sin(\pi x)\\
+\frac{c_1(q)}{4\pi^2}\left(\psi^{(2)}\left(\frac{\csc(\pi x)}{2}\right)+12\sin^2(\pi x)+8\sin^3(\pi x)-4\pi x\sin(2\pi x)\right)+\mathcal{O}((\log L)^{-2}).
\end{split}
\ee
\par From the above equation, in contrast to the previous case, we find that the equipartition of relative entropy breaks down at order $1/\log L$.
\subsection{Resolution of relative entropy between the derivative operators and the vertex operators}
In this subsection, we finally study the symmetry resolved relative entropy between two excited states, generated by a derivative operator and a vertex operator respectively. As usual, we first compute
\be\label{FVP}
\begin{split}
F^{i\p\phi,V_{\a}}_n(\m,x)&=\frac{\langle V_{\frac{\m}{2\pi}}(i\inf) V_{-\frac{\m}{2\pi}}(-i\inf)i\p\phi(t_1^-)i\p\phi(t_1^+)\prod_{k=2}^nV_{\a}(t_k^-)V_{-\a}(t_k^+)\rangle_{cy}\langle \prod_{k=1}^ni\p\phi(t_k^-)i\p\phi(t_k^+)\rangle_{cy}}
{\langle i\p\phi(t_1^-)i\p\phi(t_1^+)\prod_{k=2}^nV_{\a}(t_k^-)V_{-\a}(t_k^+)\rangle_{cy}\langle V_{\frac{\m}{2\pi}}(i\inf) V_{-\frac{\m}{2\pi}}(-i\inf)\prod_{k=1}^ni\p\phi(t_k^-)i\p\phi(t_k^+)\rangle_{cy}}\\
&=\frac{\langle V_{\frac{\m}{2\pi}}(i\inf) V_{-\frac{\m}{2\pi}}(-i\inf)i\p\phi(t_1^-)i\p\phi(t_1^+)\prod_{k=2}^nV_{\a}(t_k^-)V_{-\a}(t_k^+)\rangle_{cy}}{\langle i\p\phi(t_1^-)i\p\phi(t_1^+)\prod_{k=2}^nV_{\a}(t_k^-)V_{-\a}(t_k^+)\rangle_{cy}\langle V_{\frac{\m}{2\pi}}(i\inf) V_{-\frac{\m}{2\pi}}(-i\inf)\rangle_{cy}}[f^{i\p\phi}_{n}(\m,x)]^{-1}.
\end{split}
\ee
To calculate the complicated correlators in the above equation, it's useful to introduce
\be
\mathcal{F}_n(\m,x)=\frac{\langle V_{\frac{\m}{2\pi}}(i\inf) V_{-\frac{\m}{2\pi}}(-i\inf)i\p\phi(t_1^-)i\p\phi(t_1^+)\prod_{k=2}^nV_{\a}(t_k^-)V_{-\a}(t_k^+)\rangle_{cy}}{\langle V_{\frac{\m}{2\pi}}(i\inf) V_{-\frac{\m}{2\pi}}(-i\inf)\rangle_{cy}\langle\prod_{k=2}^nV_{\a}(t_k^-)V_{-\a}(t_k^+)\rangle_{cy}}.
\ee
Then one can rewrite $F^{i\p\phi,V_{\a}}_n(\m,x)$ as
\be
F^{i\p\phi,V_{\a}}_n(\m,x)=\frac{\mathcal{F}_n(\m,x)}{\mathcal{F}_n(0,x)}[f^{i\p\phi}_{n}(\m,x)]^{-1}
\ee
Noticing that
\be
i\partial\phi(t)=\frac{1}{\e}\frac{\p}{\p t}V_{\e}(t)\Big|_{\e=0},
\ee
we have \cite{Ruggiero:2016khg, Capizzi:2021zga}
\be
\begin{split}
\langle V_{\frac{\m}{2\pi}}(i\inf) &V_{-\frac{\m}{2\pi}}(-i\inf)i\p\phi(t_1^-)i\p\phi(t_1^+)\prod_{k=2}^nV_{\a}(t_k^-)V_{-\a}(t_k^+)\rangle_{cy}\\
&=-\frac{1}{\e^2}\frac{\p}{\p t_1^+}\frac{\p}{\p t_1^-}\langle V_{\frac{\m}{2\pi}}(i\inf) V_{-\frac{\m}{2\pi}}(-i\inf)V_{\e}(t_1^-)V_{-\e}(t_1^+)\prod_{k=2}^nV_{\a}(t_k^-)V_{-\a}(t_k^+)\rangle_{cy}\Big|_{\e=0}\\
&=\langle V_{\frac{\m}{2\pi}}(i\inf) V_{-\frac{\m}{2\pi}}(-i\inf)\prod_{k=2}^nV_{\a}(t_k^-)V_{-\a}(t_k^+)\rangle_{cy}C_{\e=0,\a}(n,x).
\end{split}
\ee
Then we can write
\be
\mathcal{F}(\m,x)=F^{I,V_{\a}}(\m,x)C_{\e=0,\a}(n,x)=e^{i\m\a x(1-\frac{1}{n})}C_{\e=0,\a}(n,x)
\ee
where
\be
C_{\e,\a}(n,x)=-\frac{1}{\e^2}\p_{t_1^+}\p_{t_1^-}\tilde{C}_{\e,\a}(n,x),
\ee
and
\be
\begin{split}
&\tilde{C}_{\e,\a}(n,x)=\langle V_{\e}(t_1^-)V_{-\e}(t_1^+)\rangle_{cy}\\
&\times \langle V_{\frac{\m}{2\pi}}(i\inf)V_{\e}(t_1^-)\rangle_{cy}\langle V_{\frac{\m}{2\pi}}(i\inf)V_{-\e}(t_1^+)\rangle_{cy}\langle V_{-\frac{\m}{2\pi}}(-i\inf)V_{\e}(t_1^-)\rangle_{cy}\langle V_{-\frac{\m}{2\pi}}(-i\inf)V_{-\e}(t_1^+)\rangle_{cy}\\
&\times\prod_{k=2}^n\langle V_{\e}(t_1^-)V_{\a}(t_k^-)\rangle_{cy}\langle V_{\e}(t_1^-)V_{-\a}(t_k^+)\rangle_{cy}\langle V_{-\e}(t_1^+)V_{\a}(t_k^-)\rangle_{cy}\langle V_{-\e}(t_1^+)V_{-\a}(t_k^+)\rangle_{cy}.
\end{split}
\ee
After taking derivatives of $\tilde{C}_{\e,\a}(n,x)$ and taking $\e=0$, one find
\be
\mathcal{F}_n(\m,x)=\frac{e^{i\m\a x(1-\frac{1}{n})}}{4}\left(\csc^2\frac{\pi x}{n}+\a^2(n\cot\pi x-\cot\frac{\pi x}{n})^2-\frac{2i\a\m}{\pi} (n\cot\pi x-\cot\frac{\pi x}{n})-\frac{\m^2}{\pi^2}\right).
\ee
It follows that
\be\label{pnparVmu}
\begin{split}
&p_n^{i\p\phi|V_{\a}}(\m)=F^{i\p\phi,V_{\a}}_n(\m,x)p_n^{i\p\phi}(\m)=\frac{\mathcal{F}_n(\m,x)}{\mathcal{F}_n(0,x)}p^I_n(\m)\\
&=\left(1-\frac{2 i\m d_n +\m^2}{\pi^2\csc^2\frac{\pi x}{n}+d_n^2}\right)e^{i\m\a x(1-\frac{1}{n})}p_n^I(\m),
\end{split}
\ee
where we have defined
\be
d_n\equiv\pi\a(n\cot\pi x-\cot\frac{\pi x}{n}).
\ee
After Fourier transformation, we obtain
\be
p_n^{i\p\phi|V_{\a}}(q)=\left(1-\frac{2d_n\D\tilde{q}+c_n(\tilde{q})\langle\D q^2\rangle^I_n}{(\pi^2\csc^2\frac{\pi x}{n}+d_n^2)\langle\D q^2\rangle^I_n}\right)p^I_n(\tilde{q}),\qquad \tilde{q}\equiv q-\a x(1-\frac{1}{n}).
\ee
The symmetry resolved R\'enyi relative entropies are given by
\be
\begin{split}
S_n(\rho_{i\p\phi}(q)\|\rho_{V_{\a}}(q))&=-\log\frac{p^{i\p\phi}(q)}{p^{V_{\a}}(q)}+\frac{1}{1-n}\log\frac{p_n^{{i\p\phi}|V_{\a}}(q)}{p^{i\p\phi}_n(q)}+S_n(\rho_{i\p\phi}\|\rho_{V_{\a}}).
\end{split}
\ee
Similar to the previous calculation, in the physical regime $\D q$ of order 1, we have
\be
\begin{split}
&S_n(\rho_{i\p\phi}(q)\|\rho_{V_{\a}}(q))=S_n(\rho_{i\p\phi}\|\rho_{V_{\a}})+\frac{2\a x\D q-\a^2x^2}{2\langle\D q^2\rangle^I}-\frac{2n\a x\D q+(1-n)\a^2 x^2}{2n^2\langle\D q^2\rangle_n^I}\\
&+\frac{c_1(q)}{\pi^2}\sin^2\pi x+\frac{1}{1-n}\left(\frac{c_n(q)}{2}b_{n,2}-\frac{2d_n\D\tilde{q}+c_n(\tilde{q})\langle\D q^2\rangle^I_n}{(\pi^2\csc^2\frac{\pi x}{n}+d_n^2)\langle\D q^2\rangle^I_n}\right)+\mathcal{O}((\log L)^{-2}).
\end{split}
\ee
The symmetry resolved relative entropy is obtained after taking the limit $n\rightarrow 1$
\be
\begin{split}
&S(\rho_{i\p\phi}(q)\|\rho_{V_{\a}}(q))=(2+\a^2)(1-\pi x\cot(\pi x))+2\log(2\sin(\pi x))+2\psi\left(\frac{\csc(\pi x)}{2}\right)+2\sin(\pi x)\\
&+\frac{c_1(q)}{4\pi^2}\left(\psi^{(2)}\left(\frac{\csc(\pi x)}{2}\right)+12\sin^2(\pi x)+8\sin^3(\pi x)-4\pi x\sin(2\pi x)\right)-\frac{\a^2x^2}{2\langle\D q^2\rangle^I}\\
&+\frac{\a\D q(2x\sin^2\pi x+(\pi\sin 2\pi x-2\pi^2x)\langle\D q^2\rangle^{I})}{\pi^2[\langle\D q^2\rangle^{I}]^2}+\mathcal{O}((\log L)^{-2}).
\end{split}
\ee
\par From the above equation we know that the equipartition of relative entropy also breaks down at order $1/\log L$.

\section{Numerical tests}\label{section5}
\begin{figure}
        \centering
        \subfloat
        {\includegraphics[width=5cm]{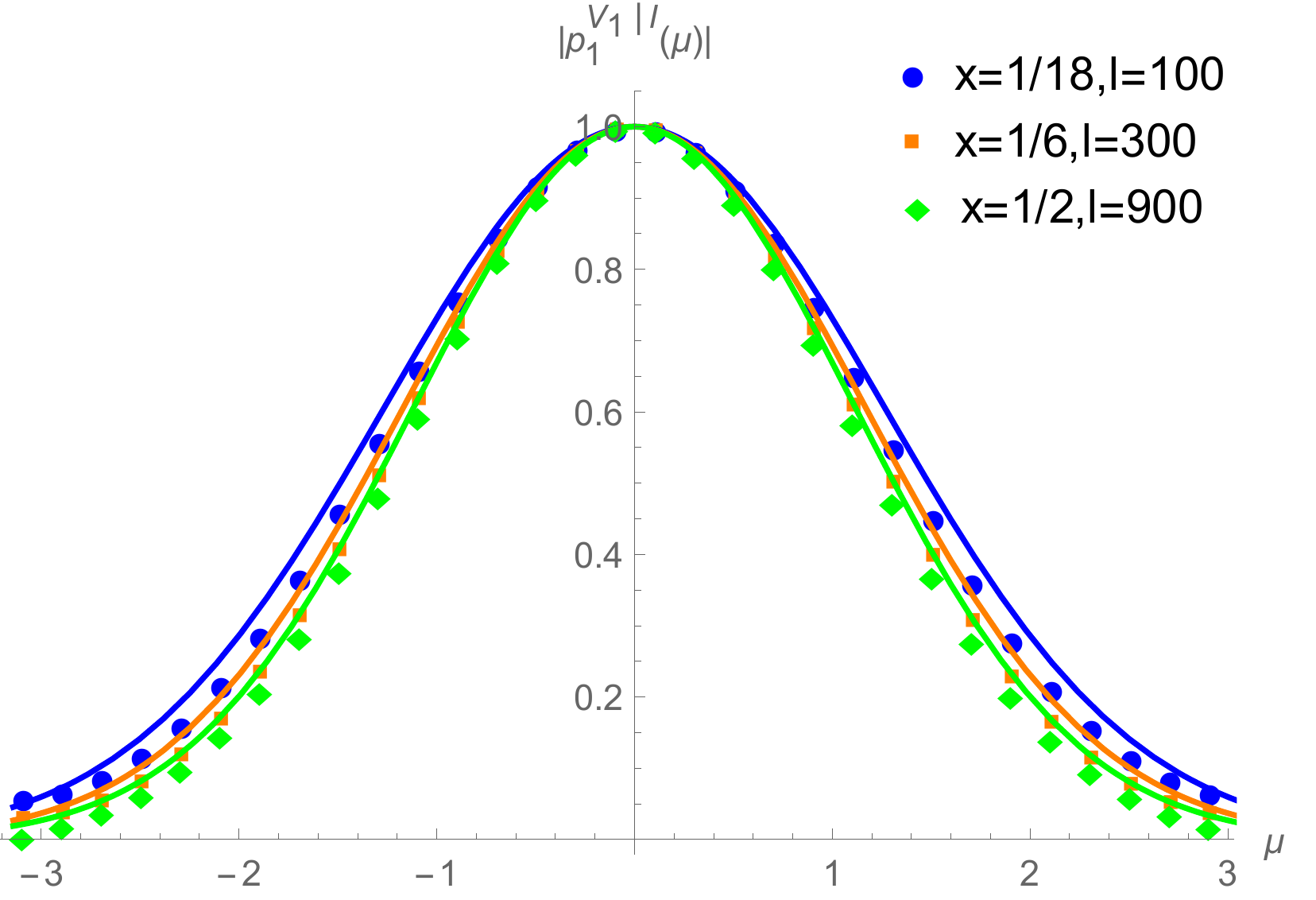}} \quad
        {\includegraphics[width=5cm]{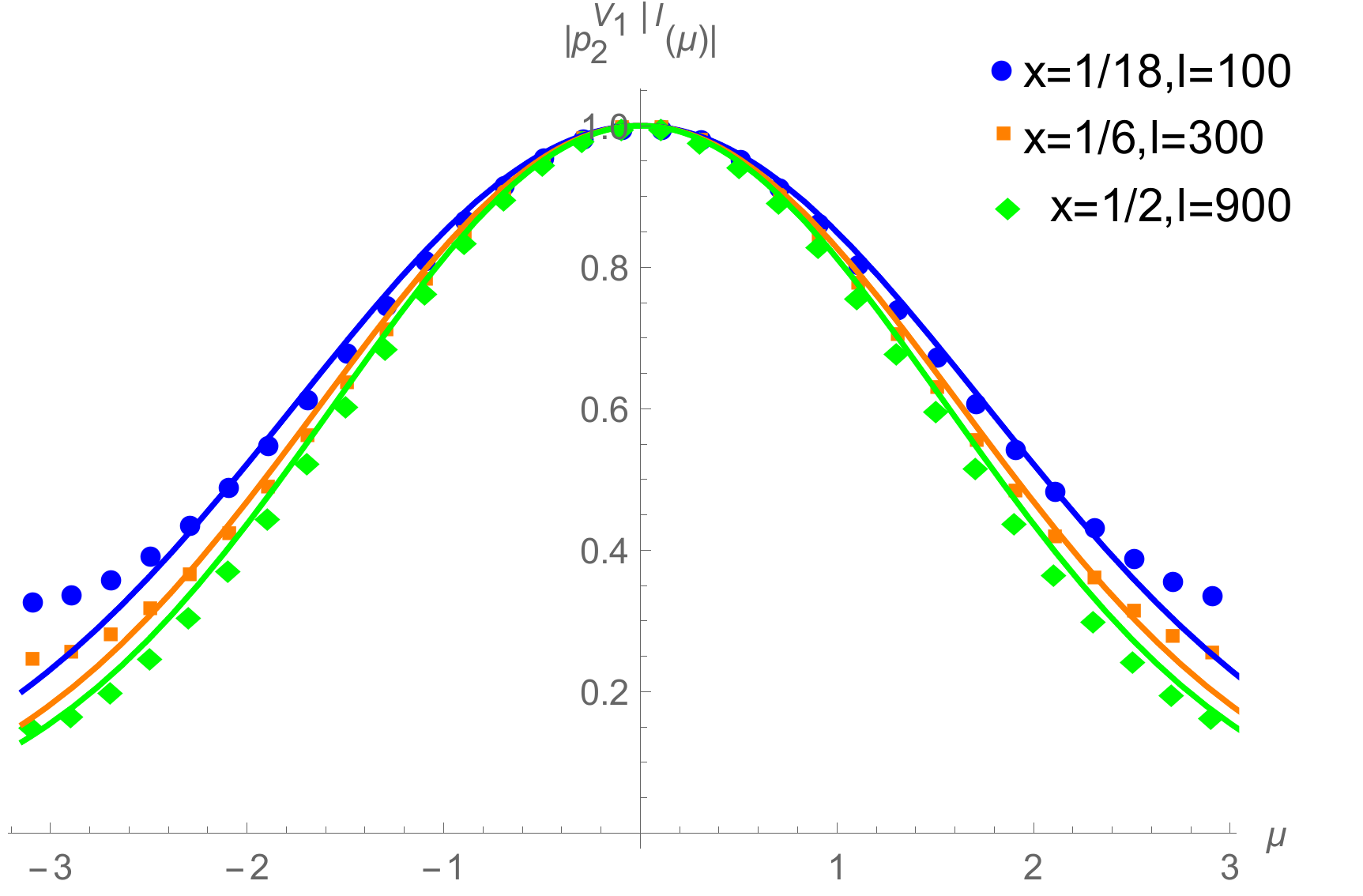}} \quad
        {\includegraphics[width=5cm]{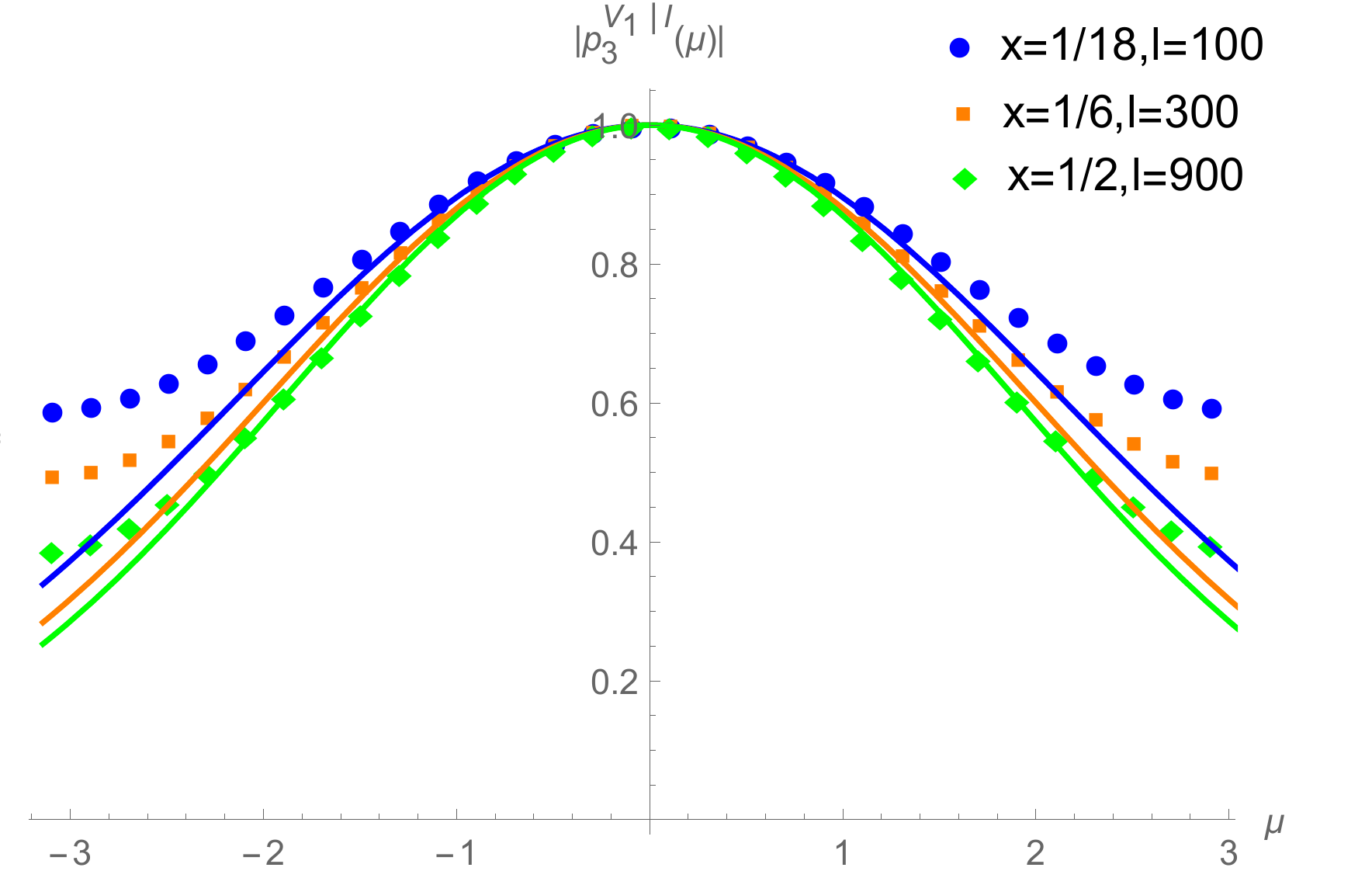}}
        \caption{Numerical data of $|p_n^{V_1|I}(\m)|$ in the XX spin chain. The full lines are the CFT predictions, eq.~(\ref{pnVImu}) with eq.~(\ref{pnImu}). Here
        we consider $n=1,2,3$ and $x=1/18,1/6,1/2$ with $L=1800$. The agreement is very well for small $\m$, but it worsens as $\m$ gets closer to $\pm\pi$ and as $n$ get larger.}
        \label{fig1}
\end{figure}

We now make some numerical tests of the universal CFT results obtained in the previous section. In this section, we will use the XX spin chain model as a concrete lattice model to check our CFT predictions. The Hamiltonian of the XX spin chain is given by
\be
H=-\frac{1}{4}\sum_{j=1}^L(\s^x_j\s^x_{j+1}+\s^y_j\s^y_{j+1}-h\s^z_j),
\ee
where $\s^{x,y,z}_j$ are the Pauli matrices acting on the $j$-th site and we impose periodic boundary conditions. For simplicity, we assume that $h=0$ and the length of chain $L$ multiples of 4.
We are only interested in the spatial bipartition of the system where subsystem $A$ is given by $l$ continuous lattice sites.
\par We did not manage to compute symmetry resolved relative entropies numerically. Instead, we numerically calculate the Fourier transformed generalized probabilities $p^{\rho|\s}_n(\m)$ for integer $n$, which are the key integrant in the computation of symmetry resolved R\'enyi relative entropies. They are defined as
\be
p^{\rho|\s}_n(\m)=\frac{\tr(\rho_A\s^{n-1}_Ae^{i\m Q_A})}{\tr(\rho_A\s^{n-1}_A)}
\ee
In the limit $L\rightarrow\inf$ with $x=l/L$ kept fixed, our numerical results should converge to the CFT computations for $p^{\rho|\s}_n(\m)$ calculated from eq.~(\ref{pnPsiPhimu}). The technical details of the numerical computation are discussed in appendix \ref{appenA}.
\par The numerical results for the function $|p_n^{V_1|I}(\m)|$  are reported in Fig.~\ref{fig1} for different $n$ and different subsystem sizes $l$, where we have used the exact results for the ground state variance given in \cite{Bonsignori:2019naz}. As shown in the figure, the agreement between numerical data and CFT prediction is excellent for small $\m$, while it gets worse for larger values of $\m$ and $n$.
\par
\begin{figure}
        \centering
        \subfloat
        {\includegraphics[width=5.2cm]{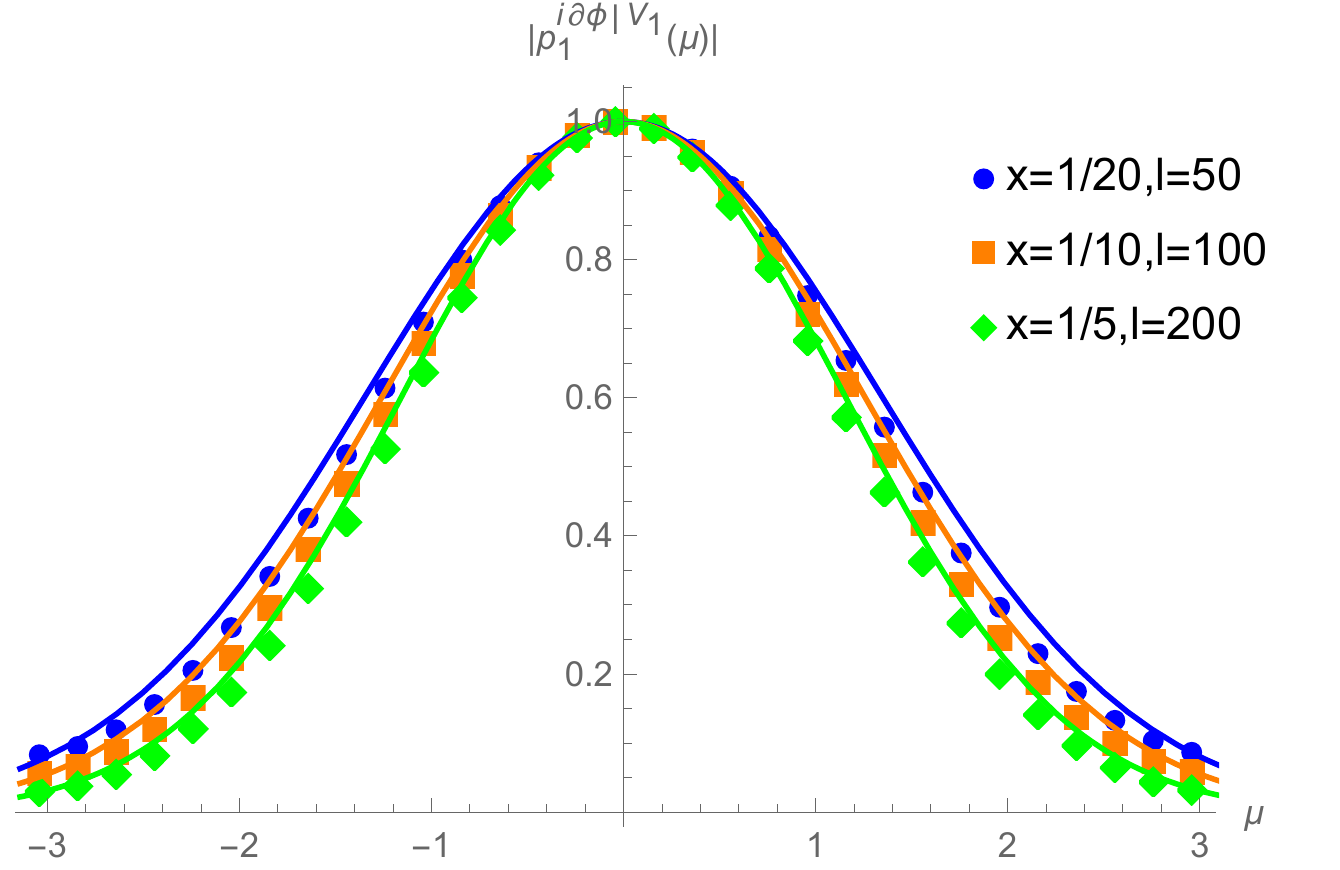}} \quad
        {\includegraphics[width=5.2cm]{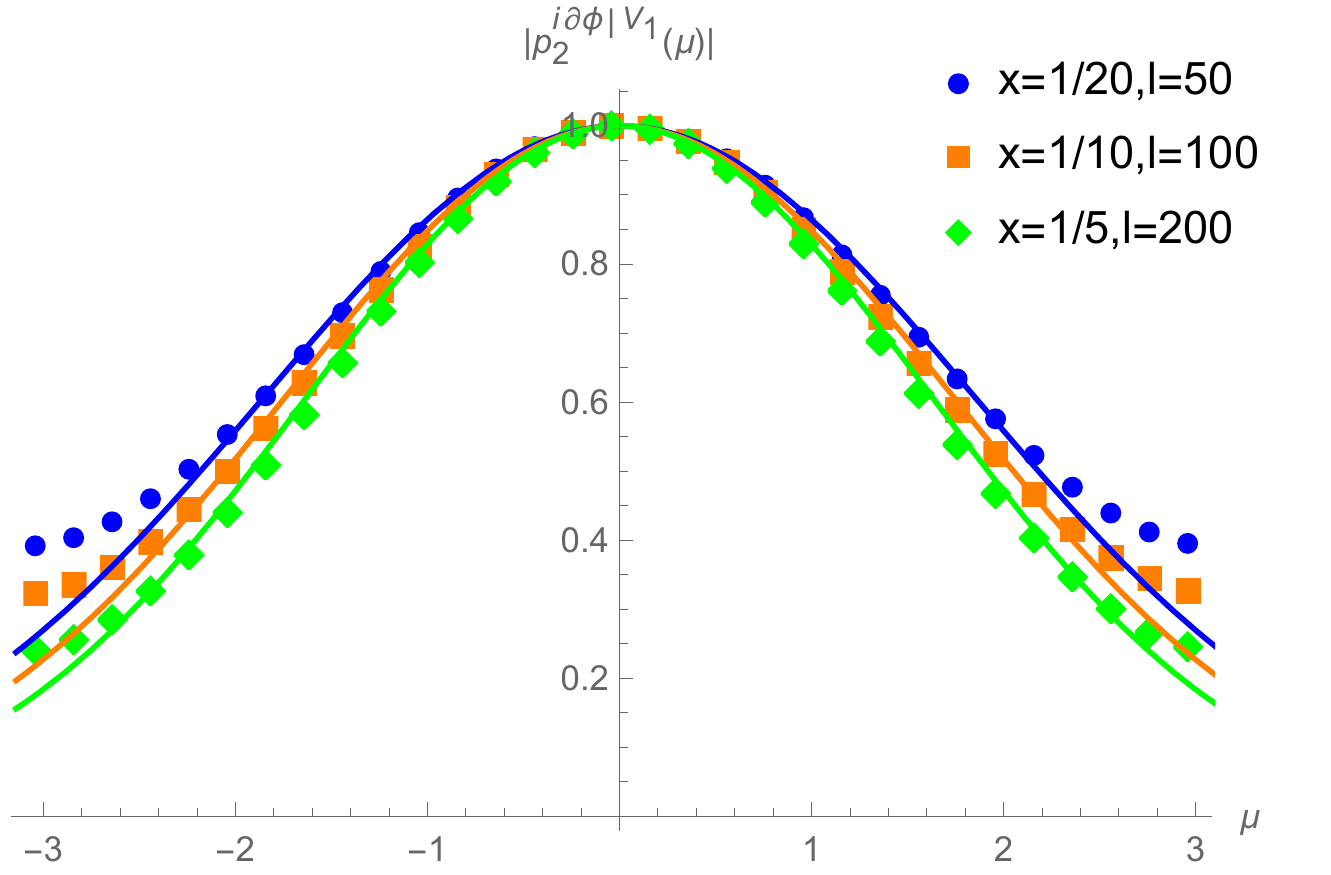}} \quad
        {\includegraphics[width=5.2cm]{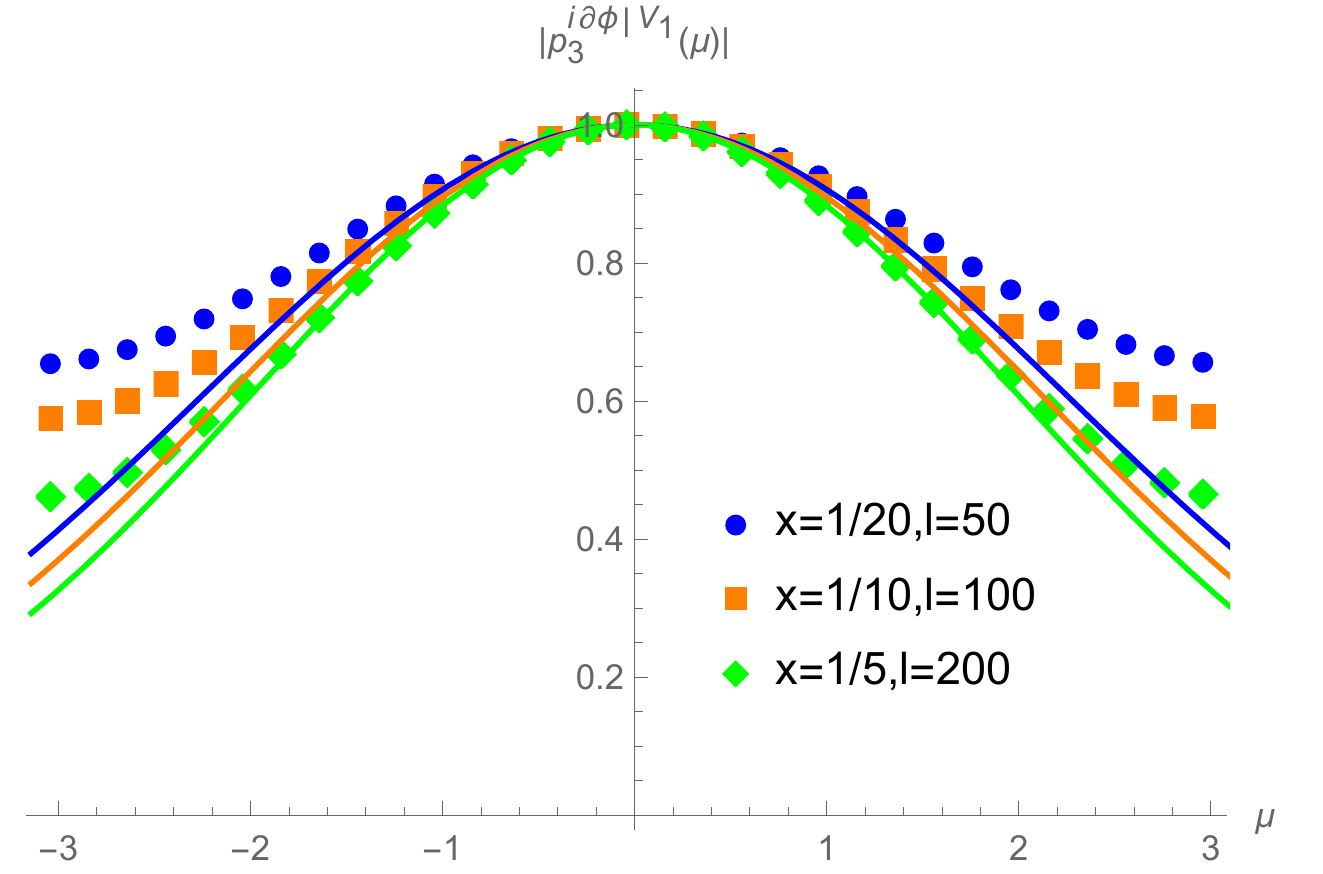}}
        \caption{Numerical data of $|p_n^{i\p\phi|V_1}(\m)|$ in the XX spin chain. The full lines are the CFT predictions, eq.~(\ref{pnparVmu}) with eq.~(\ref{pnImu}). Here
        we consider $n=1,2,3$ and $x=1/20,1/10,1/5$ with $L=1000$. The agreement is very well for small $\m$, but it worsens as $\m$ gets closer to $\pm\pi$ and as $n$ get larger.}
        \label{fig2}
\end{figure}

 In Fig.~\ref{fig2}, we report the numerical data for the quantities $|p_n^{i\p\phi|V_1}(\m)|$ for various $n$ and $l$. From eq.~(\ref{pnparVmu}), it's clear that $|p_n^{i\p\phi|V_1}(\m)|$ is non-Gaussian although it's not easy to see this from the figure. In this case, the numerical results and the CFT predictions also match very well for small values of $\m$ and $n$.
\par
\begin{figure}
        \centering
        \subfloat
        {\includegraphics[width=5.2cm]{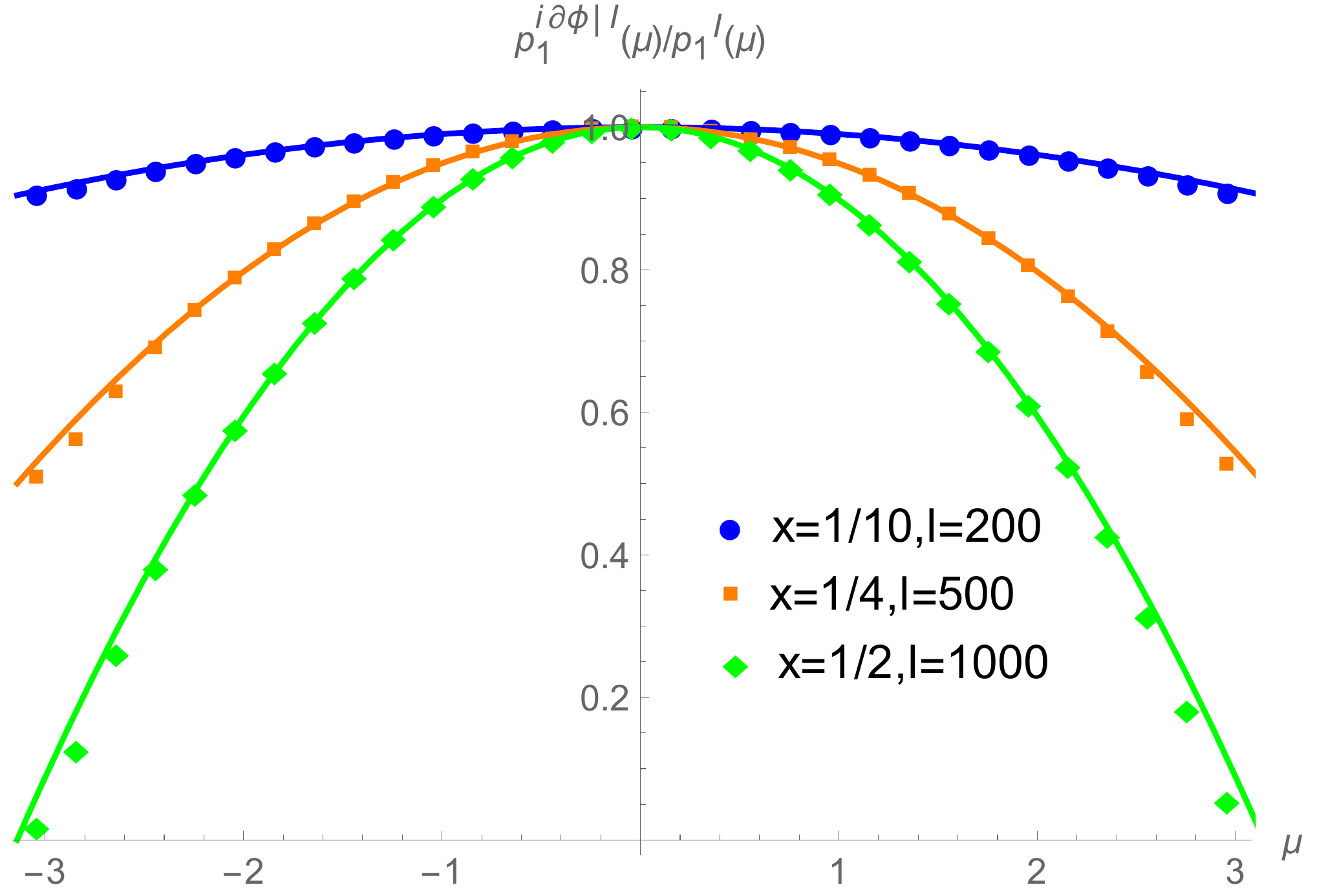}} \quad
        {\includegraphics[width=5.2cm]{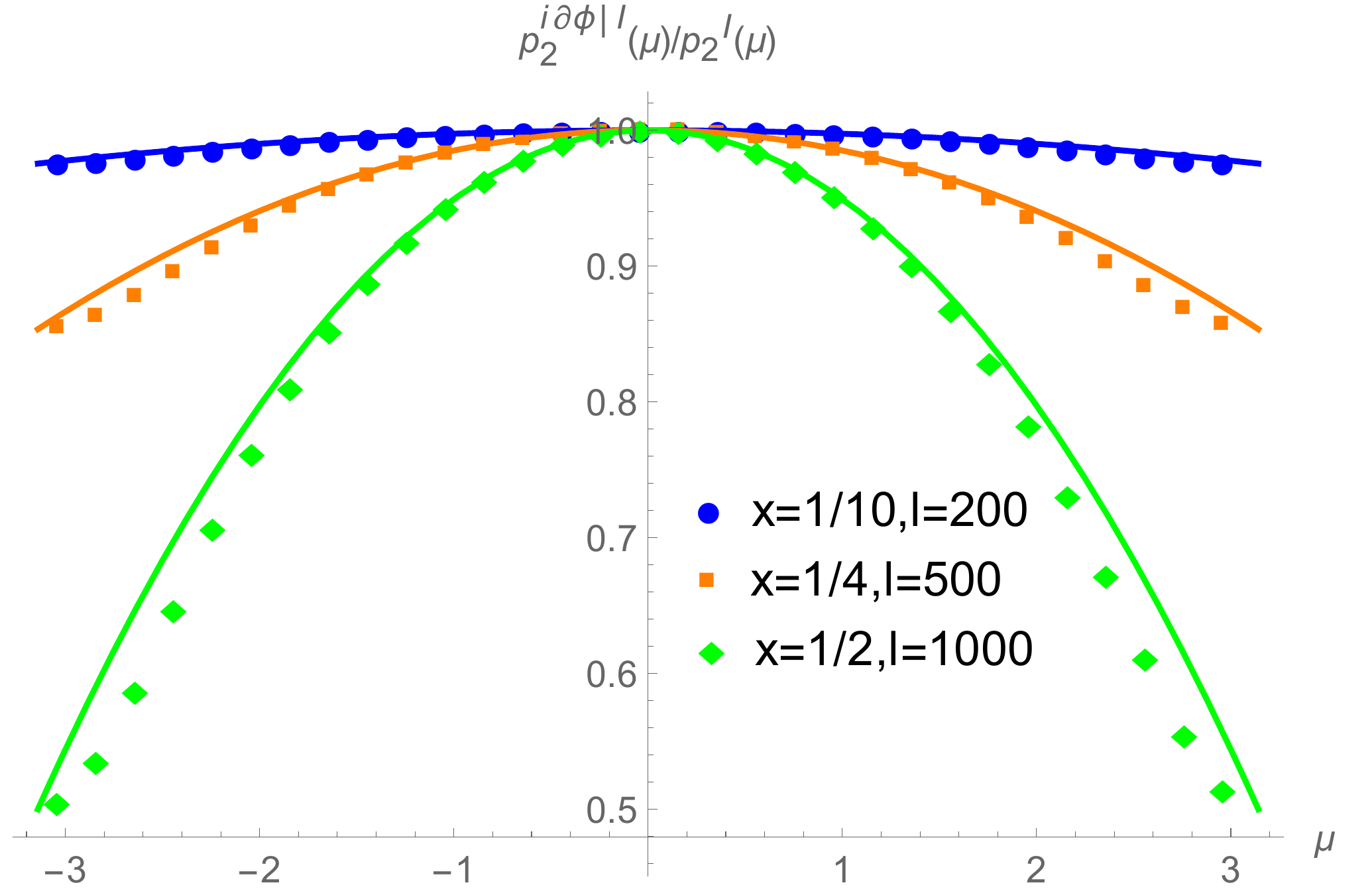}} \quad
        {\includegraphics[width=5.2cm]{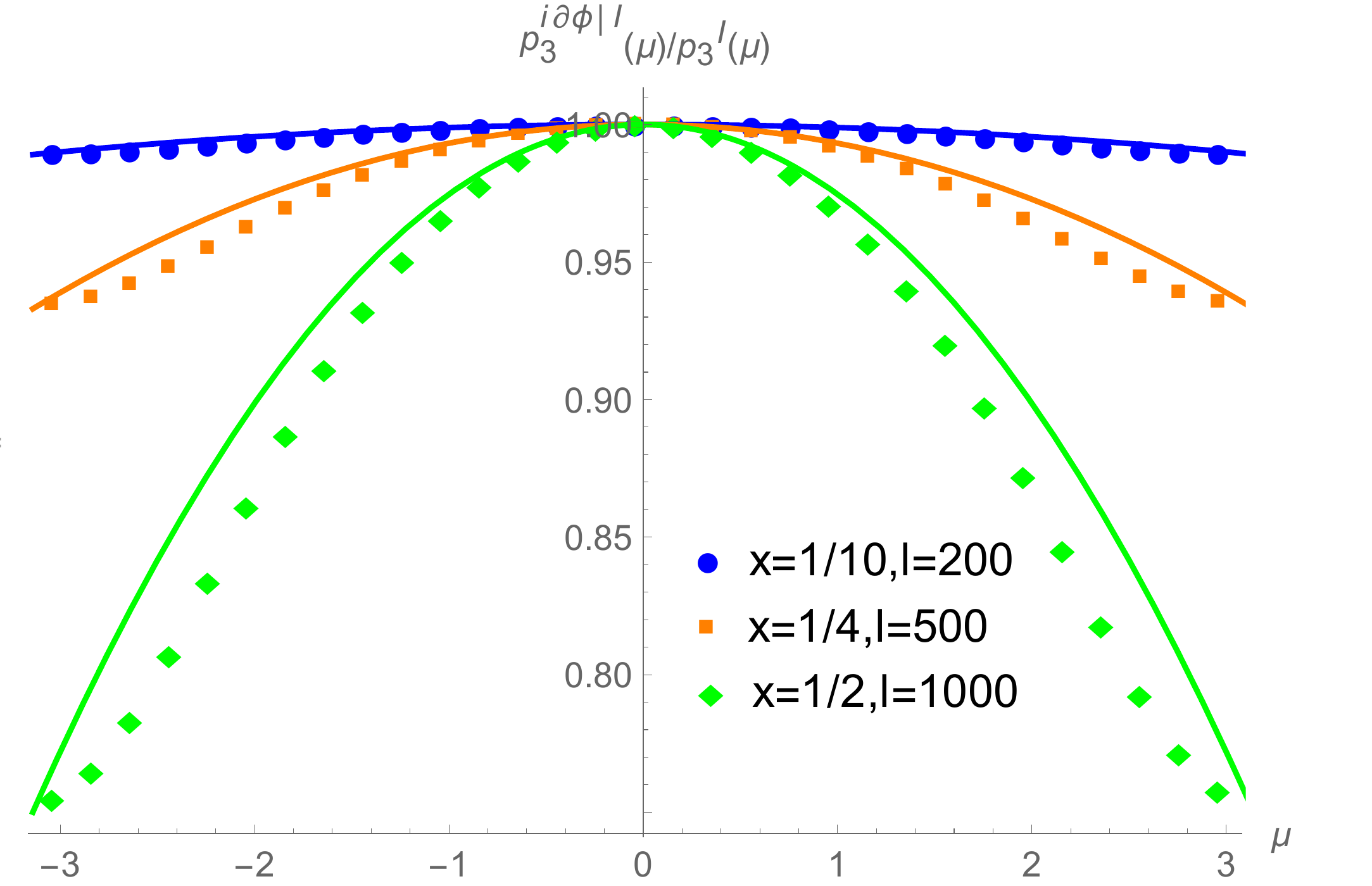}}
        \caption{Numerical data of $p_n^{i\p\phi|I}(\m)/p_n^I(\m)$ in the XX spin chain. The full lines are the CFT predictions, eq.~(\ref{pnparImu}). Here
        we consider $n=1,2,3$ and $x=1/10,1/4,1/2$ with $L=2000$. Again, the agreement is very well for small $\m$, but it worsens as $\m$ gets closer to $\pm\pi$ and as $n$ get larger.}
        \label{fig3}
\end{figure}

 In Fig.~\ref{fig3}, we show the numerical results of the ratio $p_n^{i\p\phi|I}(\m)/p_n^I(\m)$ for different $n$ and $l$. From eq.~(\ref{pnparImu}), it's easy to see that $p_n^{i\p\phi|I}(\m)$ is also non-Gaussian. This figure shows the non-Gaussian feature clearly and the agreement between the numerical data and CFT results is perfect for small values of $\m$ and $n$.
\section{Conclusion}\label{section6}
In this paper, we study the $U(1)$ symmetry resolution of relative entropies between primary states in the free massless compact boson CFT and its concrete lattice realization, the XX spin chain. We obtain various exact results from the CFT calculation using the replica method. We also carefully test our CFT predictions with the exact lattice calculations in the XX spin chain and find perfect agreements.
\par We must mention that the symmetry resolved relative entropies cannot be obtained directly by our numerical method. Instead, we just compute the Fourier transformed generalized probabilities numerically. It would be very interesting to further numerically confirm our CFT results of symmetry resolved relative entropies by other methods.
\par Several generalizations of this paper would be worth investigating. For example, one can consider the $U(1)$ symmetry resolved relative entropies in other CFTs (such as Ising and other minimal models) and the corresponding lattice models or consider the extension to Wess-Zuminon-Witten models which have non-abelian symmetries. One can also work out the symmetry resolution of other entanglement-related quantities. For example, a natural extension could be to consider the mutual information or the trace distance.
\section*{Acknowledgments}
The work was supported  by the National Natural Science Foundation of China, Grant No.\ 12005081.
\begin{appendix}
\section{Correlation matrices and RDMs in XX spin chain}\label{appenA}
The Hamiltonian of the XX spin chain is given by
\be
H=-\frac{1}{4}\sum_{j=1}^L(\s^x_j\s^x_{j+1}+\s^y_j\s^y_{j+1}-h\s^z_j),
\ee
where $\s^{x,y,z}_j$ are the Pauli matrices acting on the $j$-th site and we impose periodic boundary conditions.
After a Jordan-Wigner transformation
\be
c_j=\left(\prod_{k=1}^{j-1}\s_k^z\right)\frac{\s_j^x-i\s_j^y}{2},\qquad c_j^{\dg}=\left(\prod_{k=1}^{j-1}\s_k^z\right)\frac{\s_j^x+i\s_j^y}{2},
\ee
the spin chain Hamiltonian is mapped into a free fermion Hamiltonian on the lattice
\be\label{Hf}
H=-\frac12\sum_{j=1}^L\left[c_j^{\dg}c_{j+1}+c_{j+1}^{\dg}-2h(c_j^{\dg}c_j-\frac12)\right],
\ee
where $c_j,c^{\dg}_j$ are fermionic annihilation and creation operators, satisfying $\{c_i,c^{\dg}_j\}=\d_{ij}$. We will impose anti-periodic boundary conditions to the fermions $c_{L+1}=-c_1,c^{\dg}_{L+1}=-c^{\dg}_1$. For simplicity, we assume that $h=0$ and the length of chain $L$ multiples of 4.
The Hamiltonian eq.~(\ref{Hf}) can be diagonalized by Fourier transformation
\be
b_k=\frac{1}{\sqrt{L}}\sum_{l=1}^Lc_le^{i\phi_k l},\quad b_k^{\dg}=\frac{1}{\sqrt{L}}\sum_{l=1}^Lc_l^{\dg}e^{-i\phi_k l},\quad \phi_k=\frac{2\pi k}{L}
\ee
Then
\be
H=\sum_{k\in\O}\e_k(b_k^{\dg}b_k-\frac12),
\ee
where $\e_k=-\cos k$ and the corresponding $\O$ is
\be
\O=\{\pm\frac{1}{2},\pm\frac{3}{2},\cdots\pm\frac{L-1}{2}\}.
\ee
The eigenstates of the Hamiltonian can be characterized by a set of momenta $K$, $\ket{K}=\prod_{k\in K}b_k^{\dg}\ket{0}$. The ground state is a Fermi sea with Fermi momentum $k_F=\pi/2$ and is half-filling with fermion number $n_F=L/2$, characterized by the set of momenta: $\{\pm\frac{1}{2},\pm\frac{3}{2},\cdots\pm\frac{n_F-1}{2}\}$. Low-lying excited states are obtained by removing/adding particles in momentum space close to the Fermi surface. The correspondence between low-lying excitations in XX chain and primary excited states in CFT is described in detail in \cite{Berganza:2011mh}. This model has a $U(1)$ symmetry with the conserved charge $Q=\sum_{j=1}^Lc_j^{\dg}c_j$.

We are interested in the spatial bipartition of the system where subsystem $A$ is given by $l$ contiguous lattice sites. The reduced density matrix of the pure state $\ket{K}$ can be written as
\be
\rho_A=\det{C_A^K}\exp\left(\sum_{i,j}\log[((C_A^K)^{-1}-1)]_{ij}c_i^{\dg}c_j\right).
\ee
where the $l\times l$ matrix $[C_A^K]_{m,n}=\bra{K}c_m^{\dg}c_n\ket{K}$ is the correlation matrix restricted in $A$.
The element of $C_A^K$ is given by
\be
[C_A^K]_{mn}=\frac{1}{L}\sum_{k\in K}e^{i\phi_k(m-n)}.
\ee
It useful to introduce $2L$ Majorana modes
\be
a_{2m-1}=c_m+c_m^{\dg},\qquad a_{2m}=i(c^{\dg}_m-c_m).
\ee
For an interval with $l$ sites of the spin chain in a state $\ket{K}$, one defines the Majorana correlation matrix
\be
\langle a_ra_s\rangle_K=\d_{rs}+\G^K_{rs}
\ee
with $\G\in\mc{M}_{2N}(\mathbb{C})$:
\be\label{Gamma}
\G^K=\begin{pmatrix}
T^K_0&T^K_1&\cdots&T^K_{l-1}\\
T^K_{-1}&T^K_0&\cdots&T^K_{l-2}\\
\vdots&\vdots&\ddots&\vdots\\
T^K_{1-l}&T^K_{2-l}&\cdots&T^K_0
\end{pmatrix},
\qquad T^K_{m-n}=\begin{pmatrix}
f^K_{m-n}&g^K_{m-n}\\
-g^K_{n-m}&f^K_{m-n}
\end{pmatrix},
\ee
and
\be
\begin{split}
&f^K_{m-n}=[C^K_A]_{mn}-[C^K_A]_{nm}=\frac{1}{L}\sum_{k\in K}e^{i\phi_k(m-n)}-\frac1L\sum_{k\in K}e^{-i\phi_k(m-n)},\\
&g_{m-n}^K=-i[C^K_A]_{mn}-i[C^K_A]_{nm}+\d_{mn}=-\frac{i}{L}\sum_{k\in K}e^{-i\phi_k(m-n)}+\frac{i}{L}\sum_{k\notin K}e^{i\phi_k(m-n)}.
\end{split}
\ee
The $2^l\times 2^l$ RDM are completely determined by the correlation matrix $\G^K$.
\be
\rho_{\G_1}\rho_{\G_2}=\tr(\rho_{\G_1}\rho_{\G_2})\rho_{\G_1\times\G_2},
\ee
where
\be
\tr(\rho_{\G_1}\rho_{\G_2})=\sqrt{\left|\frac{1+\G_1\G_2}{2}\right|},
\ee
and one defines the product rule \cite{Fagotti:2010yr}
\be
\G_1\times\G_2=1-(1-\G_2)(1+\G_1\G_2)^{-1}(1-\G_1).
\ee
Now, by associativity, one can obtain the trace of the product of arbitrary number of RDMs
\be
\tr(\r_{\G_1}\r_{\G_2}\cdots)=\tr(\r_{\G_1}\r_{\G_2})\tr(\r_{\G_1\times\G_2}\cdots).
\ee
Then the above formula is rather useful to calculate the R\'enyi relative entropy.
However, when we want to know the symmetry resolution of relative entropy, we need to calculate
\be
p^{\rho|\s}_n(\m)=\frac{\tr(\rho_A\s^{n-1}_Ae^{i\m Q_A})}{\tr(\rho_A\s^{n-1}_A)}.
\ee
We can also view $e^{i\m Q_A}$ as some RDM with Majorana correlation matrix $\G^{\m}$ since
\be
\begin{split}
&e^{i\m Q_A}=e^{i\m\sum_{j=1}^lc_j^{\dg}c_j}=\prod_{j=1}^l(e^{i\m}c_j^{\dg}c_j+c_jc_j^{\dg})\\
&=(1+e^{i\m})^l\prod_{j=1}^l[p_jc_jc_j^{\dg}+(1-p_j)c_j^{\dg}c_j],
\end{split}
\ee
where $ p_j=\frac{1}{1+e^{i\m}}$. The Majorana correlation matrix $\G^\m$ have the same structure with $\G^K$ (cf.~(\ref{Gamma})) but with different block matrices
\be
f^{\m}_{m-n}=0,\quad g^{\m}_{m-n}=\frac{i(1-e^{i\m})}{e^{i\m}+1}\d_{mn}.
\ee
\end{appendix}

\bibliography{2021}
\bibliographystyle{ieeetr}
\end{document}